\documentclass[twocolumn,showpacs,aps,prd,floatfix]{revtex4}

\usepackage{graphicx}
\usepackage{dcolumn}
\usepackage{epsfig}
\usepackage{psfrag}
\usepackage{amsmath}
\newcommand{\BaBarYear}    {09}
\newcommand{\BaBarNumber}  {20}

\newcommand{\SLACPubNumber} {13785}

 \newcommand{\BaBarType}      {PUB}  % Journal publication

\input babarsym   

\RequirePackage{xspace}

\newcommand{\pvec}{{\bf p}}
\newcommand{\DE}{\ensuremath{\Delta E}}

\newcommand{\xf}{\ensuremath{{\cal F}}}
\newcommand{\hel}{\ensuremath{{\cal H}}}
\newcommand{\thetaT}{\ensuremath{\theta_{\rm T}}}
\newcommand{\costhr}{\ensuremath{\cos\thetaT}}

\newcommand\etal{{\it et al.}}
\newcommand{\half}{\ensuremath{\frac{1}{2}}}

\newcommand{\bfig}{\begin{figure}[htbpc!]}
\newcommand{\efig}{\end{figure}}
\newcommand\bef{\begin{figure}}
\newcommand\edf{\end{figure}}
\newcommand\beq{\begin{equation}}
\newcommand\eeq{\end{equation}}
\newcommand\bear{\begin{array}}
\newcommand\enar{\end{array}}
\newcommand\beqa{\begin{eqnarray}}
\newcommand\eeqa{\end{eqnarray}}
\newcommand\ben{\begin{enumerate}}
\newcommand\een{\end{enumerate}}

\newcommand{\UfourS}{\ensuremath{\Upsilon(4S)}}

\def\Bz      {\ensuremath{B^0}\xspace}
\def\Bbar    {\kern 0.18em\overline{\kern -0.18em B}{}\xspace}
\def\BB{\ensuremath{B\Bbar}\xspace} 
\def\Dbar    {\kern 0.18em\overline{\kern -0.18em D}{}\xspace}

\newcommand{\Kuno}{\ensuremath{K_1(1270)}}
\newcommand{\Kunop}{\ensuremath{K_1(1400)}}
\newcommand{\Ku}{\ensuremath{K_1}}

\begin{document}

\begin{flushleft}
\babar-\BaBarType-\BaBarYear/\BaBarNumber \\
SLAC-PUB-\SLACPubNumber \\
%arXiv:\LANLNumber
\end{flushleft}

\title{\boldmath Measurement of branching fractions of $B$ decays to
$K_1(1270)\pi$ and $K_1(1400)\pi$ and determination of the CKM angle $\alpha$ 
from $B^0 \ra a_1(1260)^{\pm} \pi^{\mp}$}

%% author list as of 02-Jun-2009 (489 authors)
%
\author{B.~Aubert}
\author{Y.~Karyotakis}
\author{J.~P.~Lees}
\author{V.~Poireau}
\author{E.~Prencipe}
\author{X.~Prudent}
\author{V.~Tisserand}
\affiliation{Laboratoire d'Annecy-le-Vieux de Physique des Particules (LAPP), Universit\'e de Savoie, CNRS/IN2P3,  F-74941 Annecy-Le-Vieux, France}
\author{J.~Garra~Tico}
\author{E.~Grauges}
\affiliation{Universitat de Barcelona, Facultat de Fisica, Departament ECM, E-08028 Barcelona, Spain }
\author{M.~Martinelli$^{ab}$}
\author{A.~Palano$^{ab}$ }
\author{M.~Pappagallo$^{ab}$ }
\affiliation{INFN Sezione di Bari$^{a}$; Dipartimento di Fisica, Universit\`a di Bari$^{b}$, I-70126 Bari, Italy }
% ...
\collaboration{The \babar\ Collaboration}
\noaffiliation

\begin{abstract}
We report measurements of the branching fractions of neutral and 
charged $B$ meson decays to final states containing a 
$K_1(1270)$ or $K_1(1400)$ 
meson and a charged pion. 
The data, collected with the {\slshape B\kern-0.1em{\smaller A}\kern-0.1em
    B\kern-0.1em{\smaller A\kern-0.2em R}} detector at the SLAC National
Accelerator Laboratory, 
correspond to 454 million $B\kern 0.18em\overline{\kern -0.18em B}{}$ 
pairs produced in
$e^+e^-$ annihilation. 
We measure the branching fractions
${\cal B}(B^0\rightarrow K_1(1270)^{+}\pi^-+K_1(1400)^{+}\pi^-)= 3.1^{+0.8}_{-0.7} \times 10^{-5}$ and 
${\cal B}(B^+\rightarrow K_1(1270)^{0}\pi^++K_1(1400)^{0}\pi^+)= 2.9^{+2.9}_{-1.7} \times 10^{-5}$ ($<8.2\times 10^{-5}$ at 90\% confidence level),
where the errors are statistical and systematic combined.
The $B^0$ decay mode is observed with a significance 
of $7.5\sigma$, while a significance of $3.2\sigma$ is 
obtained for the $B^+$ decay mode.
Based on these results, we estimate the 
weak phase 
$\alpha=(79 \pm 7 \pm 11)^{\circ}$ 
from the time dependent $CP$ asymmetries in 
$B^0 \rightarrow a_1(1260)^{\pm} \pi^{\mp}$ decays.

\end{abstract}

% PACS, the Physics and Astronomy Classification Scheme.

\maketitle

\section{Introduction}
\label{sec:intro}

$B$ meson decays to final states containing  
an axial-vector meson ($A$) and a pseudoscalar meson ($P$)
have been studied both 
theoretically and experimentally.
Theoretical predictions for the branching fractions (BFs)
of these decays
have been calculated assuming a na\"{\i}ve 
factorization hypothesis \cite{LAPORTA,CALDERON} and QCD
factorization~\cite{CHENG}. 
These decay modes are expected to occur with BFs of order $10^{-6}$.
Branching fractions of $B$ meson decays 
with an $a_1(1260)$ or $b_1(1235)$ meson plus a   
pion or a kaon in the final state have recently been 
measured \cite{APDECAYSA1,APDECAYSB1}. 

The \babar\ Collaboration has measured \CP-violating
asymmetries in $B^0 \ra a_1(1260)^{\pm} \pi^{\mp}$ decays and
determined an effective value $\alpha_{\rm eff}$~\cite{ALPHA} for 
the phase angle $\alpha$ of the Cabibbo-Kobayashi-Maskawa (CKM) 
quark-mixing matrix~\cite{CKM}. 
In the absence of penguin (loop) contributions in these decay modes, 
$\alpha_{\rm eff}$ coincides with $\alpha$.

The $\Delta S = 1$ decays we examine here are particularly sensitive to
the presence of penguin amplitudes because their CKM
couplings are larger than the corresponding $\Delta S = 0$ penguin amplitudes.
Thus measurements of the decay rates of the $\Delta S = 1$ transitions
involving the same SU(3) flavor multiplet as $a_1(1260)$ provide constraints on
$\Delta\alpha=\alpha_{\rm eff} -\alpha$~\cite{BOUNDS}.
Similar SU(3)-based approaches have been proposed for the extraction of 
$\alpha$ in the $\pi^+\pi^-$~\cite{PIPIPH}, 
$\rho^{\pm}\pi^{\mp}$~\cite{BOUNDS}, and 
$\rho^+\rho^-$ channels~\cite{RHORHOPH,RHORHOEXP}.

The rates of $B \rightarrow K_{1A} \pi$ decays, 
where the $K_{1A}$ meson is the SU(3) partner of $a_1(1260)$ and a 
nearly equal admixture of the \Kuno\ and \Kunop\ 
resonances~\cite{PDG},
can be derived from the rates of $B \rightarrow K_1(1270) \pi$ and 
$B \rightarrow K_1(1400) \pi$ decays.
For $B^0 \ra K_1(1400)^+ \pi^-$~\cite{CHARGE}
and $B^+ \ra K_1(1400)^0 \pi^+$
decays there exist experimental upper limits at the $90\%$ confidence 
level (C.L.) of $1.1\times10^{-3}$ and $2.6\times10^{-3}$,
respectively~\cite{ARGUS}.
In the following, \Ku\ will be used to indicate both $K_1(1270)$ and $K_1(1400)$ mesons.

The production of $K_1$ mesons in $B$ decays has been 
previously observed in the $B \to J/\psi K_1$, $B \to K_1\gamma$, 
and $B \to K_1\phi$ decay channels~\cite{BTOK1}. 
Here we present measurements of the 
$B^0 \ra K_1^+ \pi^-$ 
and $B^+ \ra K_1^0 \pi^+$ branching 
fractions and estimate the weak phase $\alpha$ from the 
measurement of the time dependent $CP$ asymmetries in 
$B^0 \ra a_1(1260)^{\pm} \pi^{\mp}$ decays and 
the branching fractions of SU(3) related modes.

This paper is organized as follows.
In Sec.\ \ref{detector} we describe the dataset and the detector.
In Sec.\ \ref{sec:Model} we introduce the $K-$matrix formalism used 
for the parameterization of the $K_1$ resonances.
Section \ref{selection} is devoted to a discussion of the 
reconstruction and selection of the $B$ candidates.
In Sec.\ \ref{sec:MLfit} we describe the maximum likelihood fit for 
the signal branching fractions and the likelihood scan over the 
parameters that characterize the production of the $K_1$ system.
In Sec.\ \ref{sec:syst} we discuss the systematic uncertainties.
In Sec.\ \ref{sec:results} we present the experimental results.
Finally, in Sec.~\ref{sec:alpha}, we use the experimental results 
to extract bounds on $|\Delta \alpha|$.

\section{The BaBar detector and dataset}
\label{detector}
The results presented in this paper are based on data collected with the
\babar\ detector at the \pep2\ asymmetric-energy \epem\ storage ring,
operating at the SLAC National Accelerator Laboratory. At \pep2, 9.0\,\gev\
electrons collide with 3.1\,\gev\ positrons to yield a center-of-mass 
(CM) energy
of $\sqrt{s}=10.58$\,\gev, which corresponds to the mass of the \FourS\
resonance.  The asymmetric energies result in a boost from the laboratory to
the %\epem\ 
CM frame of $\beta\gamma\approx 0.56$.  We
analyze the final \babar\ dataset collected at the \FourS\ resonance,
corresponding to an integrated luminosity of 413~fb$^{-1}$ and $N_{\BB} = (454.3 \pm 5.0)\times 10^6$ produced \BB\ pairs. 

A detailed description of the \babar\ detector can be found
elsewhere~\cite{BABARNIM}.  Surrounding the interaction point is a
five-layer double-sided silicon vertex tracker (SVT) that 
provides precision measurements near the collision point of
charged particle tracks in the planes transverse
to and along the beam direction. A 40-layer drift chamber 
surrounds 
the SVT.  Both of these tracking devices operate in the 1.5\,T magnetic
field of a superconducting solenoid to provide measurements of the
momenta of charged particles. 
Charged hadron identification is achieved through
measurements of particle energy loss in the tracking system and the
Cherenkov angle obtained from a detector of internally reflected Cherenkov
light. 
A CsI(Tl) electromagnetic calorimeter 
provides photon detection and electron identification. 
Finally, the instrumented flux return (IFR) of the magnet
allows discrimination of muons from pions and detection of \KL\ mesons.  For
the first 214$\invfb$ of data, the IFR was composed of a resistive plate
chamber system.  For the most recent 199$\invfb$ of data, a portion
of the resistive plate chamber system has been replaced by limited streamer
tubes~\cite{lsta}.

We use a GEANT4-based Monte Carlo (MC) simulation to model the response of the
detector \cite{geant}, taking into account the varying accelerator and 
detector conditions. We generate large samples of signal and background 
for the modes considered in the analysis.

\section{Signal model}
\label{sec:Model}

In this analysis the signal is characterized by two nearby resonances, 
$K_1(1270)$ and $K_1(1400)$, which have the same quantum numbers, 
$I(J^P)=1/2(1^+)$, and
decay predominantly to the same $K\pi\pi$ final state.
The world's largest sample of $K_1(1270)$ and $K_1(1400)$ events was 
collected by the ACCMOR Collaboration with the WA3 experiment~\cite{WA3}. 
The WA3 fixed target experiment accumulated data 
from the reaction $K^-p\rightarrow K^-\pi^+\pi^-p$ with an incident kaon energy 
of $63$\,\gev.
These data were analyzed using a 
two-resonance, six-channel $K$-matrix model \cite{KMATRIX}
to describe the resonant $K\pi\pi$ system. 
We base our parameterization of the $K_1$ resonances produced 
in $B$ decays on a model derived from the $K$-matrix 
description of the scattering amplitudes in Ref.\ \cite{WA3}. 
In Sec.~\ref{formalism} we briefly outline the 
$K$-matrix formalism, which is then applied 
in Sec.~\ref{fitwa3} to fit the ACCMOR data in order to 
determine the 
parameters describing the diffractive production of $K_1$ mesons 
and their decay. 
In Sec.~\ref{bdecays} we explain how we use the extracted values
of the decay parameters  and describe our model for
$K_1$ production in $B\rightarrow (K\pi\pi)\pi$ decays.

\subsection{$K$-matrix formalism}
\label{formalism}
Following the analysis of the ACCMOR Collaboration, 
the $K\pi\pi$ system is described by a $K$-matrix model 
comprising six channels, 
$1=(K^*(892)\pi)_S$, $2=\rho K$, $3=K_0^*(1430) \pi$, $4=f_0(1370) K$, 
$5=(K^*(892) \pi)_D$, $6=\omega K$.
We identify each channel by the intermediate resonance and bachelor 
particle, where the bachelor particle is
the $\pi$ or $K$ produced directly from the $K_1$ decay. For the 
$K^*(892) \pi$ channels the subscript refers to the angular momentum.

We parameterize the production amplitude for each
channel in the reaction $K^-p\rightarrow (K^-\pi^+\pi^-) p$ as
\begin{equation}
F_i=e^{i \delta_i}\sum_j(\mathbf{1}-i\mathbf{K}{\boldsymbol \rho})_{ij}^{-1}P_j,
\label{Fvector}
\end{equation}
where the index $i$ (and similarly $j$) represents the $i^{th}$ channel.
The elements of the diagonal phase space matrix 
\mbox{\boldmath$\rho$}$(M)$
for the decay chain 
\begin{equation}
\label{decaychain}
K_1 \rightarrow 3 + 4, ~3 \rightarrow 5 + 6, 
\end{equation}
are approximated with the form 
\begin{equation}
\rho_{ij}(M)=\frac{2\delta_{ij}}{M}\sqrt{\frac{2m^*m_4}{m^*+m_4}(M-m^*-m_4+i\Delta)},
\end{equation}
where $M$ is the $K\pi\pi$ invariant mass,  $m_4$ is the mass of the 
bachelor particle
$4$, and $m^*$ ($\Delta$) is the pole mass (half width) 
of the intermediate resonance state $3$~\cite{PHSP}. 
In Eq. (\ref{Fvector}), the $\delta_i$ parameters are offset phases with 
respect to the $(K^*(892) \pi)_S$ channel ($\delta_1 \equiv 0$).
The $6\times 6$ $K$-matrix has the following form:
\begin{equation}
K_{ij} = \frac{f_{ai}f_{aj}}{M_a-M}+\frac{f_{bi}f_{bj}}{M_b-M},
\end{equation}
where the labels $a$ and $b$ refer to $K_1(1400)$ and $K_1(1270)$, respectively.
The decay  constants $f_{ai}$, $f_{bi}$  and the $K$-matrix poles
$M_{a}$  and $M_{b}$ are real. 
The production vector $\mathbf{P}$ consists of a background term $\mathbf{D}$ 
\cite{DECK} and a direct production term $\mathbf{R}$
\begin{equation}
\mathbf{P} = (\mathbf{1}+\tau\mathbf{K})\mathbf{D}+\mathbf{R},
\label{Pvector}
\end{equation}
where $\tau$ is a constant.

The background amplitudes are parameterized by
\begin{equation}
D_i = D_{i0}\frac{e^{i\phi_i}}{M^2-M_K^2}
\end{equation}
for all channels but $(K^*(892) \pi)_D$ and $\omega K$.
For the $(K^*(892) \pi)_D$ channel we set $D_5=0$ as in the 
ACCMOR analysis~\cite{WA3}. The parameters for the $\omega K$ channel are 
not fitted, as described later in this Section, and we set $D_6=0$. 
The results are not sensitive to this choice for the value of $D_6$.

$\mathbf{R}$ is given by
\begin{equation}
R_i = \frac{f_{pa}f_{ai}}{M_a-M}+\frac{f_{pb}f_{bi}}{M_b-M},
\end{equation}
where $f_{pa}$ and $f_{pb}$ represent the amplitude for producing the 
states $K_1(1400)$ and $K_1(1270)$, respectively, and are complex numbers.
We assume $f_{pa}$ to be real.

P-wave ($\ell=1$) and D-wave ($\ell=2$) centrifugal barrier factors
are included in the $K_1$ decay couplings $f_{ai}$ and $f_{bi}$ 
and background amplitudes $D_{i0}$, and are given by:
\begin{equation}
  B_{i}(M)= \left[ \frac{q_i(M)^2 R^2}{1+q_i(M)^2 R^2} \right]^{\ell/2},
\end{equation}
where $q_i$ is the breakup momentum in channel $i$.
Typical values for the interaction radius squared $R^2$ are in the range
$5 < R^2 < 100\,\gev^{-2}$~\cite{THREEPI} and the value 
$R^2=25~\gev^{-2}$ is used.

The physical resonances $K_1(1270)$ and $K_1(1400)$ are mixtures of
the two SU(3) octet states $K_{1A}$ and $K_{1B}$:
\begin{eqnarray}
|K_1(1400)\rangle &=& |K_{1A}\rangle \cos\theta + |K_{1B}\rangle \sin\theta, \\
|K_1(1270)\rangle &=& - |K_{1A}\rangle \sin\theta + |K_{1B}\rangle \cos\theta.  
\end{eqnarray}
Assuming that SU(3) violation manifests itself only in the mixing,
we impose the following relations~\cite{WA3}:
\begin{eqnarray}
f_{a1}=&\frac{1}{2}\gamma_+\cos{\theta} + \sqrt{\frac{9}{20}}\gamma_-
\sin{\theta},\\ 
f_{b1}=&-\frac{1}{2}\gamma_+\sin{\theta} + \sqrt{\frac{9}{20}}\gamma_-
\cos{\theta},\\ 
f_{a2}=&\frac{1}{2}\gamma_+ \cos{\theta} - \sqrt{\frac{9}{20}}\gamma_-
\sin{\theta},\\ 
f_{b2}=&-\frac{1}{2}\gamma_+ \sin{\theta} - \sqrt{\frac{9}{20}}\gamma_-
\cos{\theta},
\end{eqnarray}
where $\gamma_+$ and $\gamma_-$ are the couplings of the SU(3) octet
states to the $(K^*(892)\pi)_S$ and $\rho K$ channels:
$\langle(K^*(892)\pi)_S|K_{1A}\rangle~=~\frac{1}{2}\gamma_+~=~\langle\rho K |K_{1A}\rangle$ and $\langle\rho K|K_{1B}\rangle = -\sqrt{\frac{9}{20}}\gamma_-~=~-\langle (K^*(892) \pi)_S|K_{1B}\rangle$.
The couplings for the $\omega K$ channel 
are fixed to $1/\sqrt{3}$ of the $\rho K$ 
couplings, 
as follows from the quark model \cite{WA3}.

\subsection{Fit to WA3 data}
\label{fitwa3}
\begin{figure}[!htb]
  \begin{center}
    \includegraphics[width=8.6cm]{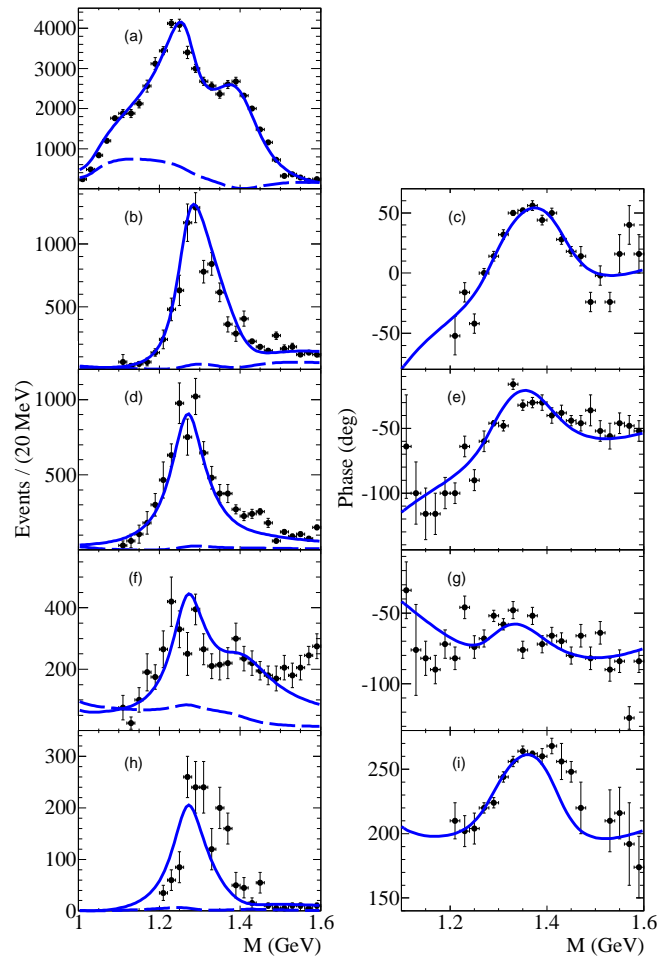}
 \caption{
   Results of the fit to the $0\le |t'| \le 0.05\,\gev^2$ WA3 data.
   Intensity (left) and phase 
   relative to the $(K^*(892)\pi)_S$ amplitude (right) 
   for the (a) $(K^*(892)\pi)_S$, 
   (b, c) $\rho K$, (d, e) $K_0^*(1430) \pi$, (f, g) $f_0(1370) K$, and 
   (h, i) $(K^*(892) \pi)_D$ 
   channels. The points represent the data, the solid lines 
   the total fit function, and the dashed lines the contribution 
   from the background.
 } 
 \label{fig:daumfit}
  \end{center}
\end{figure}

Only some of the $K$-matrix parameters extracted in the ACCMOR analysis 
have been reported in the literature \cite{WA3}. In particular, the 
results for most of the decay couplings $f_{ai}$ and $f_{bi}$ are not
available.
The ACCMOR Collaboration performed a partial-wave analysis of the 
WA3 data. The original WA3 paper \cite{WA3}
provides the results of the partial-wave analysis of the  $K\pi\pi$
system in the form of plots 
for the intensity in  the $(K^*(892)\pi)_S$, $\rho K$, $K_0^*(1430) \pi$, 
$f_0(1370) K$, and $(K^*(892) \pi)_D$ channels, together with the phases of the 
corresponding amplitudes, measured relative to the $(K^*(892)\pi)_S$ amplitude.
The $\omega K$ data were not analyzed.
In order to obtain an estimate of the parameters that enter
the $K$-matrix model, we perform a $\chi^2$ fit of this model to the $0\le |t'| \le 0.05~\gev^2$ WA3
data for the intensity of the $m=0$ $K\pi\pi$ channels and the relative phases. 
Here $|t'|$ is the four momentum transfer 
squared with respect to the recoiling proton in the reaction 
$K^-p\rightarrow K^-\pi^+\pi^-p$, and $m$ denotes the magnetic substate of 
the $K\pi\pi$ system.
Since the results of the analysis performed by the ACCMOR Collaboration are 
not sensitive to the choice of the value for the constant $\tau$ in 
Eq. (\ref{Pvector}), we set $\tau=0$. 
We seek solutions corresponding to positive values of the $\gamma_{\pm}$ 
parameters, as found in the ACCMOR analysis \cite{WA3}. 
The data sample consists of 215 bins. The results of this fit are displayed 
in Fig.~\ref{fig:daumfit} and show
a good qualitative agreement with the results obtained by the ACCMOR 
Collaboration \cite{WA3}.
We obtain $\chi^2=855$, with 26 free parameters, 
while the ACCMOR Collaboration obtained $\chi^2=529$. 
Although neither fit is formally a good one, the 
model succeeds in reproducing the relevant features of the 
data.

\subsection{Model for $K_1$ production in $B$ decays}
\label{bdecays}

We apply the above formalism to the parameterization of the signal 
component for the production of $K_1$ resonances in $B$ decays. 
The propagation of the uncertainties in the $K-$matrix description
of the ACCMOR data to the model for $K_1$ production in $B$ decays is a 
source of systematic uncertainty and is taken into account as described in 
Sec.~\ref{sec:syst}.

In order to parameterize the signal component for the analysis of $B$ decays,
we set the background amplitudes $\mathbf{D}$, whose contribution 
should be small in the non-diffractive case, to 0. The backgrounds arising from 
resonant and non-resonant $B$ decays to the $(K\pi\pi)\pi$ final state 
are taken into account by separate components in the fit, as described in 
Sec.~\ref{sec:MLfit}.
The parameters of $\mathbf{K}$ and the offset 
phases $\delta_i$ are assumed to be independent from the production process 
and are fixed to the values extracted from the fit to WA3 data 
(Table~\ref{tab:daumfit}). 
\begin{table}[h]
\begin{center}
\begin{tabular}{cc}
\hline\hline 
Parameter    & Fitted value\\
\hline 
$M_a        $&$ 1.40 \pm 0.02 $\\
$M_b        $&$ 1.16 \pm 0.02 $\\
\hline
$\theta     $&$ 72^{\circ} \pm 3^{\circ}$\\
$\gamma_+   $&$ 0.75 \pm 0.03 $\\
$\gamma_-   $&$ 0.44 \pm 0.03 $\\
$f_{a3}     $&$ 0.02 \pm 0.03 $\\
$f_{b3}     $&$ 0.32 \pm 0.01 $\\
$f_{a4}     $&$-0.08 \pm 0.02 $\\
$f_{b4}     $&$ 0.16 \pm 0.01$\\
$f_{a5}     $&$ 0.06 \pm 0.01$\\
$f_{b5}     $&$ 0.21 \pm 0.04$\\
\hline
$\delta_{2}$&$-31^{\circ} \pm 1^{\circ}$\\
$\delta_{3}$&$ 82^{\circ} \pm 2^{\circ}$\\
$\delta_{4}$&$ 78^{\circ} \pm 4^{\circ}$\\
$\delta_{5}$&$ 20^{\circ} \pm 9^{\circ}$\\
\hline\hline
\end{tabular}
\end{center}
\caption{Parameters for the $K$-matrix model used in the analysis of $B$ decays.}
\label{tab:daumfit}
\end{table}
Finally, we express the production couplings $f_{pa}$ and $f_{pb}$ in terms
of two real production parameters ${{\boldsymbol \zeta}=(\vartheta,\phi)}$:
$f_{pa}\equiv \cos\vartheta$, $f_{pb}\equiv \sin\vartheta e^{i\phi}$, where
$\vartheta \in [0,\pi/2]$, $\phi \in [0,2\pi]$. 
In this parameterization, $\tan\vartheta$ represents the magnitude of the 
production constant for the $K_1(1270)$ resonance relative to that for the 
$K_1(1400)$ resonance, while $\phi$ is the relative phase.

The dependence of the selection efficiencies and of the distribution
of the discriminating variables (described in Sec.~\ref{sec:MLfit}) 
on the production parameters 
${\boldsymbol \zeta}$ are derived from Monte Carlo studies.
For given values of ${\boldsymbol \zeta}$, signal MC samples for $B$ 
decays to the $(K\pi\pi)\pi$ final states are generated by weighting the
$(K\pi\pi)\pi$ population according to the amplitude
$\sum_{i\neq \omega K}\langle K\pi\pi| i \rangle F_i$, where the term 
$\langle K\pi\pi| i \rangle$
consists of a factor describing the angular distribution 
of the $K\pi\pi$ system resulting from $K_1$ decay, an amplitude 
for the resonant $\pi\pi$ and $K\pi$ systems, and isospin factors,
and is calculated using the formalism described in Refs.~\cite{WA3,HERNDON}. 
The $\omega K$ channel is excluded from the sum, since 
the $\omega \to \pi^+\pi^-$ branching fraction is only 
$(1.53^{+0.11}_{-0.13})\,\%$, compared to the branching fraction 
$(89.2\pm 0.7)\,\%$ of the dominant decay 
$\omega \to \pi^+\pi^-\pi^0$~\cite{PDG}. Most of the 
$K_1\to \omega K$ decays therefore result in a different final state than the 
simulated one. We account for the $K_1\to \omega K$ transitions with a 
correction to the overall efficiency.
\begin{figure}[!htb]
  \begin{center}
    \includegraphics[width=3in]{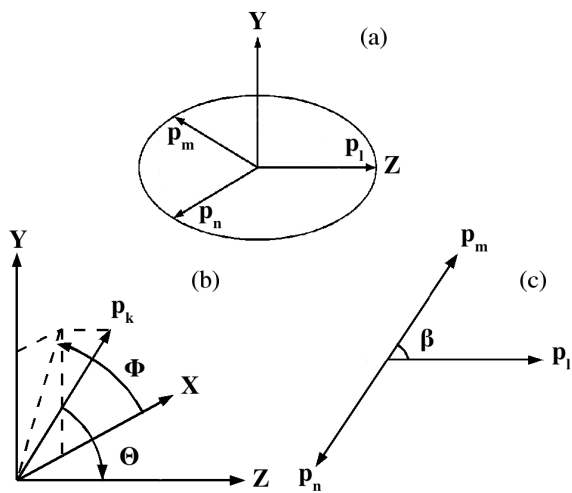}
 \caption{
Definition of (a) the coordinate axes in the $K_1$ rest
frame, (b) the angles $\Theta$ and $\Phi$ in the
$K_1$ rest frame, and (c) the angle $\beta$
in the rest frame of the $X_{s,d}$ intermediate 
resonance.
 } 
 \label{fig:refframe}
  \end{center}
\end{figure}
In Fig.~\ref{fig:refframe} we show the reference frame chosen 
to evaluate the distributions of the products of $B\to K_1\pi$ decays,
where $K_1$ decays proceed through the intermediate resonances 
$X_s=\{K^*(892),K_0^*(1430)\}$ or $X_d=\{\rho,f_0(1370),\omega\}$.
Final state particles are labeled with a subscript $\{k,l,m,n\}$, 
according to the following scheme:
$B^0\rightarrow K_1^+ \pi^-_k$, $K_1^+ \rightarrow X_s^0 \pi^+_l$
$X_s^0 \rightarrow K^+_m \pi^-_n$ or $B^0\rightarrow K_1^+ \pi^-_k$,
$K_1^+ \rightarrow X_d^0 K^+_l$, $X_d^0 \rightarrow \pi^+_m \pi^-_n$ 
for neutral $B$ meson decays, 
and 
$B^+\rightarrow K_1^0 \pi^+_k$, $K_1^0 \rightarrow X_s^+ \pi^-_l$,
$X_s^+ \rightarrow K^0_m \pi^+_n$ or $B^+\rightarrow K_1^0 \pi^+_k$, 
$K_1^0 \rightarrow X_d^0 K^0_l$, $X_d^0 \rightarrow \pi^+_m \pi^-_n$
for charged $B$ meson decays.
The angular distribution for the $K_1$ 
system produced in $B$ decays can be expressed in terms of three independent 
angles ($\Theta$, $\beta$, $\Phi$).
In the $K_1$ rest frame, we define the $Y$ axis as the normal to the decay 
plane of the $K_1$, and orient the $Z$ axis along the momentum of $l$ 
(Fig.~\ref{fig:refframe}a). $\Theta$ and $\Phi$ are then the polar and 
azimuthal angles of the momentum of $k$, respectively, in the $K_1$ rest frame 
(Fig.~\ref{fig:refframe}b).
We define $\beta$ as the polar angle of the flight direction of $m$ 
relative to the 
direction of the momentum of $l$ (Fig.~\ref{fig:refframe}c).
The resulting angular parts of the transition amplitudes 
for $S$-, $P$-, and $D$-wave 
decays of the $K_1$ axial vector ($J^P=1^+$) mesons with scalar ($J^P=0^+$) and
vector ($J^P=1^-$) intermediate resonances $X_{s,d}$ are given by:
\begin{align}
A_{S} &= \sqrt{\frac{3}{8\pi}} \left(\cos\Theta \cos\beta +
  \sin\Theta \sin\beta \cos\Phi \right) \\
A_{P} &= \sqrt{\frac{3}{8\pi}} \cos\Theta \\
A_{D} &= \sqrt{\frac{3}{16\pi}} \left(-2 \cos\Theta \cos\beta +
  \sin\Theta \sin\beta \cos\Phi \right).
\end{align}

For the $\pi\pi$ and $K\pi$ resonances, the following 
$\ell$-wave Breit-Wigner parameterization is used~\cite{HERNDON}:
\begin{equation}
BW(m)= (\pi)^{-1/2} \frac{\left[m_0 \Gamma(m)\right]^{1/2}}
{\left(m_0^2 - m^2\right)-i m_0 \Gamma(m)}
\end{equation}
with
\begin{equation}
\Gamma(m)=\Gamma(m_0)\frac{m_0}{m}\left[ \frac{q(m)}{q(m_0)} \right]^{2\ell +1} 
\left[\frac{1+R^2 q^2(m_0)}{1+R^2 q^2(m)}\right]^{\ell},
\end{equation}
where $m_0$ is the nominal mass of the resonance, $\Gamma(m)$
is the mass-dependent width, $\Gamma(m_0)$ is the nominal width of the 
resonance, $q$ is the breakup momentum of the 
resonance into the two-particle final state, and $R^2=25$ $\gev^{-2}$.
The $K_0^*(1430)$ and $f_0(1370)$ amplitudes are also 
parameterized as Breit-Wigner functions.
For the $K_0^*(1430)$ we assume a mass of $1.250$\,\gev\ and 
a width of $0.600$\,\gev~\cite{WA3}, while for the $f_0(1370)$ we 
use a mass of $1.256$\,\gev and a width of $0.400$\,\gev~\cite{KELLY}.
This parameterization is varied in Sec.~\ref{sec:syst} and a 
systematic uncertainty evaluated.

\section{Event reconstruction and selection}
\label{selection}

The $B^0 \rightarrow K_1^+ \pi^-$ candidates are reconstructed 
in the $K_1^+ \rightarrow K^+\pi^+\pi^-$  decay mode by means
of a vertex fit of all combinations of 
four charged tracks having a zero net charge.
Similarly we reconstruct $B^+ \rightarrow K_1^0 \pi^+$ candidates, with
$K_1^0 \rightarrow K_S^0 \pi^+\pi^-$,
by combining $K_S^0$ candidates with three charged tracks. 
We require the reconstructed mass $m_{K\pi\pi}$ to lie in the 
range $\left[1.1,1.8\right]$\,\gev.
Charged particles are identified as either pions or kaons,
and must not be consistent with the electron, muon or proton 
hypotheses.
The $K_S^0$ candidates are reconstructed from pairs of oppositely-charged 
pions with an invariant mass in the range $[486,510]$\,\mev , 
whose decay vertex is required to be displaced from the $K_1$ vertex by 
at least $3$ standard deviations. 

The reconstructed $B$ candidates are characterized by two almost  
uncorrelated variables, the energy-substituted mass
\begin{eqnarray}
\mes &\equiv& \sqrt{(\half s + \pvec_0\cdot\pvec_B)^2/E_0^2 - \pvec_B^2}
\end{eqnarray}
and the energy difference
\begin{eqnarray}
\DE &\equiv& E_B^*-\half\sqrt{s},
\end{eqnarray}
where $(E_0,\pvec_0)$ and $(E_B,\pvec_B)$ are the laboratory four-momenta of
the \UfourS\ and the $B$ candidate, respectively, and the asterisk
denotes the CM frame. We require
$5.25<\mes<5.29\,\gev$ and $|\DE|<0.15\,\gev$.
For correctly reconstructed $B$ candidates, 
the distribution of \mes\ peaks at the $B$-meson mass and 
\DE\ at zero.

Background events arise primarily from random combinations of particles
in continuum $\epem\ra\qqbar$ events ($q=u,d,s,c$). 
We also consider cross feed from other \B\ meson decay modes
than those in the signal. 

To separate continuum from \BB\ events we use variables that characterize
the event shape.
We define the angle \thetaT\ between the thrust axis \cite{thrust} of the 
$B$ candidate in the \UfourS\ frame
and that of the charged tracks and neutral calorimeter clusters in the
rest of the event.  
The distribution of $|\costhr|$ is sharply peaked near 1 for \qqbar\ jet
pairs and nearly uniform for \B-meson decays.  We require
$|\costhr|<0.8$.
We construct a Fisher discriminant \xf\  
from a linear combination of
four topological variables: the monomials $L_0=\sum_ip_i^*$ and 
$L_2=\sum_ip_i^*|\cos \theta_i^*|^2$, $|\cos\theta_C^*|$ and $|\cos\theta_B^*|$~\cite{FISHER}.
Here, $p_i^*$ and $\theta_i^*$ are the CM momentum and the angle of the 
remaining tracks and clusters in the event with respect to the $B$ 
candidate thrust axis.
$\theta_C^*$ and $\theta_B^*$ are the CM polar angles of the 
$B$-candidate thrust axis and $B$-momentum vector, respectively,
relative to the beam axis.
In order to improve the accuracy in the determination of the event shape 
variables, we require a minimum of $5$ tracks in each event. 

Background from $B$ decays to final states containing charm or charmonium
mesons is suppressed by means of vetos.
A signal candidate is rejected if it shares at least 
one track with a $B$ candidate reconstructed in the
$B^0 \ra D^-\pi^+$, $B^0 \ra D^{*-} \pi^+$, $B^+ \ra \bar D^0 \pi^+$,
or $B^+ \ra \bar D^{*0} \pi^+$ decay modes, where the $D$ meson
in the final states decays hadronically.
A signal candidate is also discarded if 
any $\pi^+ \pi^-$ combination consisting of the primary pion 
from the $B$ decay together with an oppositely charged pion from the 
$K_1$ decay has an invariant mass consistent with the
$\ccbar$ mesons $\chi_{c0}(1P)$ or $\chi_{c1}(1P)$ decaying to 
a pair of oppositely charged pions, or $J/\psi$ and $\psi(2S)$ decaying to 
muons where the muons are misidentified as pions.

We define \hel\ as the cosine of the angle between the direction of the
primary pion from the $B$ decay and the normal to the plane defined by the 
\Ku\ daughter momenta in the \Ku\ rest frame. We require $|\hel| < 0.95$ to 
reduce background from $B \rightarrow V \pi$ decay modes, where $V$ is
a vector meson decaying to $K\pi\pi$, such as $K^*(1410)$ or $K^*(1680)$.

The average number of candidates in events containing at least 
one candidate is $1.2$.
In events with multiple candidates, we select the candidate 
with the highest $\chi^2$ probability of the $B$ vertex fit.

We classify the events according to the invariant masses
of the $\pi^+ \pi^-$  and $K^+ \pi^-$ ($K_S^0 \pi^+$)
systems in the 
$K_1^+$ ($K_1^0$) decay for $B^0$ ($B^+$) candidates:
events that satisfy the requirement 
$0.846 < m_{K\pi} < 0.946$ \gev\ belong to class 1 (``$K^*$ band''); 
events not included in class 1 for which $0.500 <m_{\pi\pi} < 0.800$ \gev\
belong to class 2 (``$\rho$ band''); all other events are rejected.
The fractions of selected signal events in class 1 and class 2 range from 
33\% to 73\% and from 16\% to 49\%, respectively, depending on the 
production parameters $\boldsymbol{\zeta}$. About 11\% to 19\% of 
the signal events are rejected at this stage. For combinatorial background,
the fractions of selected events in class 1 and class 2 are 
22\% and 39\%, respectively, while 39\% of the events are rejected.

The signal reconstruction and selection efficiencies depend on the 
production parameters $\boldsymbol{\zeta}$. For $B^0$ modes these efficiencies 
range from
5 to 12\% and from 3 to 8\% for events in class 1 and class 2, 
respectively.
For $B^+$ modes the corresponding values are 4-9\% and
2-7\%.

\section{Maximum likelihood fit}
\label{sec:MLfit}

We use an unbinned,  extended maximum-likelihood (ML) fit to extract the
event yields $n_{s,r}$ and the parameters of the probability density function
(PDF) ${\cal P}_{s,r}$. 
The subscript $r=\{1,2\}$ corresponds to one of the
resonance band classes defined in Sec.\ \ref{selection}.
The index $s$ represents the event categories used in our fit model. 
For the analysis of $B^0$ modes, these are 
\begin{enumerate}
\item signal, 
\item combinatorial background,
\item $B^0 \ra K^{*}(1410)^+ \pi^-$,
\item $B^0 \ra K^{*}(892)^0\pi^+\pi^- + \rho^0 K^+ \pi^-$, 
\item $B^0 \ra  a_1(1260)^{\pm} \pi^{\mp}$,  and
\item $B^0 \ra D^-_{K\pi\pi} \pi^+$.
\end{enumerate}
For $B^+$ modes, these are 
\begin{enumerate}
\item signal, 
\item combinatorial background,
\item $B^+ \ra K^{*}(1410)^0 \pi^+$,
\item $B^+ \ra K^{*}(892)^+\pi^+\pi^- + \rho^0 K_S^0 \pi^+$, and 
\item $B^+ \ra K^{*}(892)^+ \rho^0$.
\end{enumerate}

The likelihood ${\cal L}_{e, r}$ for a candidate $e$ 
to belong to class $r$ is defined as 
\begin{equation}
{\cal L}_{e, r} = \sum_{s}n_{s,r}\, {\cal P}_{s,r}({\boldsymbol
{\rm x}}_e;~{\boldsymbol \zeta},~{\boldsymbol \xi}),
\end{equation}
where the PDFs are formed using the set of observables
${\boldsymbol{\rm x}}_e$~$=\{\Delta E$, $\mes$, $\xf$, $m_{K\pi\pi}$, $|\hel|$\}
and the dependence on the production parameters {\boldmath$\zeta$} is
relevant only for the signal PDF.
{\boldmath$\xi$} represents all other PDF parameters.

In the definition of ${\cal L}_{e, r}$ the yields of the signal category
for the two classes are expressed as a function of the signal branching
fraction $ {\cal B}$ as
$n_{1,1}={\cal B} \times N_{\BB} \times
\epsilon_1$({\boldmath$\zeta$}) and $n_{1,2}={\cal B} \times N_{\BB}
\times \epsilon_2$({\boldmath$\zeta$}), where  the total selection
efficiency $\epsilon_r$({\boldmath$\zeta$}) includes the daughter
branching fractions and the reconstruction efficiency obtained from MC
samples as a function of the production parameters.

For the $B^0$ modes we perform a negative log-likelihood scan with respect to 
$\vartheta$ and $\phi$. Although the events in class $r=2$ are characterized by 
a smaller signal-to-background ratio with respect to the events in
class $r=1$, MC studies show that including these events in the fit for 
the $B^0$ modes helps to resolve ambiguities in the determination of 
$\phi$ in cases where a signal is observed. 
At each point of the scan, a simultaneous 
fit to the event classes $r={1,2}$ is performed.

For the $B^+$ modes, simulations show that, due to a 
less favorable signal-to-background ratio and increased 
background from $B$ decays, we are not sensitive to
$\phi$ over a wide range of possible values of the signal BF.
We therefore assume $\phi=\pi$ and restrict the scan to $\vartheta$.
At each point of the scan, we perform a fit to the events in
class $r=1$ only. The choice $\phi=\pi$ minimizes the variations 
in the fit results associated with differences between the $m_{K\pi\pi}$
PDFs for different values of $\phi$. This source of systematic 
uncertainty is accounted for as described in Sec.~\ref{sec:syst}. 
The variations in the efficiency $\epsilon_1$ as a function of $\phi$ 
for a given $\vartheta$ can be as large as $30\,\%$, and are taken 
into account in deriving the branching fraction results 
as discussed in Sec.~\ref{sec:results}. 

The fitted samples consist of 23167 events ($B^0$ modes, class $1$),
38005 events ($B^0$ modes, class $2$), and 
9630 events ($B^+$ modes, class $1$).

The signal branching fractions are free parameters in the fit.
The yields for event categories $s=5,6$ ($B^0$ modes) and $s=5$ 
($B^+$ modes) are fixed to the values estimated from MC simulated data 
and based on their previously measured branching fractions \cite{PDG,HFAG}. 
The yields for the other background components are determined from 
the fit. 
The PDF parameters for combinatorial background are left free to vary in
the fit, while those for the other event categories are fixed to the 
values extracted from MC samples.

The signal and background PDFs are constructed as products of PDFs 
describing the distribution of each observable. 
The assumption of negligible correlations in the selected data samples
among the discriminating variables has been tested with MC samples. 
The PDFs for $\Delta E$ and $\mes$ of the categories  1, 3, 4, and 5
are each parameterized 
as a sum of a Gaussian function to describe the core of 
each distribution, plus an empirical function determined from MC simulated 
data to account for the tails of each distribution.
For the combinatorial background we use a first degree Chebyshev
polynomial for $\Delta E$ and an empirical phase-space function~\cite{ARGUSMES}
for $\mes$:
\begin{equation}
f(x)\propto x\sqrt{1-x^2}\exp\left[-\xi_1(1-x^2)\right],
\end{equation}
where $x\equiv 2\mes/\sqrt{s}$ and $\xi_1$ is a parameter that is determined 
from the fit.
The combinatorial background PDF is found to describe well both the
dominant quark-antiquark background and the background from random 
combinations of $B$ tracks.

For all categories the $\xf$ distribution is well described by 
a Gaussian function with different widths to the left and right of the 
mean. A second Gaussian function with a larger width accounts for a small
tail in the distribution and prevents the background probability 
from becoming too small in the signal $\xf$ region.

The $m_{K\pi\pi}$ distribution for signal depends on {\boldmath$\zeta$}.
To each point of the {\boldmath$\zeta$} scan, we therefore associate 
a different nonparametric template, modeled upon signal MC samples 
reweighted according to the corresponding values of the production 
parameters $\vartheta$ and $\phi$.
Production of $K^{*}(1410)$ and $a_1(1260)$ resonances occurs 
in $B$ background and is taken into account in the 
$m_{K\pi\pi}$ and $|{\cal H}|$ PDFs.
For all components, the PDFs for $|{\cal H}|$ are parameterized with 
polynomials.

We use large control samples to verify the 
$m_{ES}$, $\Delta E$, and $\xf$ PDF shapes,
which are initially 
determined from MC samples. 
We use the $B^0\to D^-\pi^+$ decay with $D^-\to K^+\pi^-\pi^-$, 
and the $B^+\to \bar{D}^0\pi^-$ decay with $D^0\to K^0_S\pi^+\pi^-$, 
which have similar topology to the signal $B^0$ and $B^+$ modes, 
respectively.
We select these samples by applying loose requirements on 
$m_{ES}$ and $\Delta E$, and requiring for the $D$ candidate mass
$1848<m_{D^-}<1890\,\mev$ and $1843<m_{D^0}<1885\,\mev$.
The selection requirements on the $B$ and $D$ daughters are 
very similar to those of our signal modes.
These selection criteria are applied both to the data and to the MC events.
There is good agreement between data and MC samples:
the deviations in the 
means of the distributions are about $0.5\,\mev$ for $m_{ES}$,
$3\,\mev$ for $\Delta E$, and negligible for $\xf$.

\section{Systematic uncertainties}
\label{sec:syst}

The main sources of systematic uncertainties are summarized in
Table~\ref{tab:systtab}. 
For the branching fractions, the errors that affect the result only 
through efficiencies are called ``multiplicative'' 
and given in percentage. All other errors are labeled 
``additive'' and expressed in units of $10^{-6}$.

We repeat the fit by varying the PDF parameters {\boldmath$\xi$},
within their uncertainties, 
that are not left floating in the fit.
%--------------------------------------------------------------------%
The signal PDF model excludes fake combinations originating from
misreconstructed signal events.
Potential biases due to the presence of fake combinations, 
or other imperfections in the signal PDF model, are estimated with MC
simulated data.
We also account for possible bias introduced by the finite 
resolution of the ($\vartheta,\phi$) likelihood scan.
%--------------------------------------------------------------------%
A systematic error is evaluated by varying the $K_1(1270)$ and $K_1(1400)$ 
mass poles and $K-$matrix parameters 
in the signal model, the parameterization of the intermediate 
resonances in $K_1$ decay, and the offset phases $\delta_i$.
%--------------------------------------------------------------------%
We test the stability of the fit results against variations in 
the selection of the ``$K^*$'' and ``$\rho$ bands,'' and
evaluate a corresponding systematic error. 
%--------------------------------------------------------------------%
An additional systematic uncertainty originates from potential peaking 
\BB\ background, including $B \ra K_2^{*}(1430) \pi$ and  
$B \ra K^{*}(1680) \pi$, and is evaluated by introducing
the corresponding components in the definition of the likelihood 
and repeating the fit with their yields fixed to 
values estimated from the available experimental information~\cite{PDG}.
%--------------------------------------------------------------------%
We vary the yields of the $B^0\ra  a_1(1260)^{\pm} \pi^{\mp}$ and  
$B^0 \ra D_{K^+ \pi^- \pi^-}^-\pi^+$ (for the $B^0$ modes) 
and $B^+ \ra K^{*+} \rho^0$ (for the $B^+$ modes) 
event categories by their uncertainties and take the resulting 
change in results as a systematic error.
%--------------------------------------------------------------------%
For $B^+$ modes, we introduce an additional systematic uncertainty 
to account for the variations of the $\phi$ parameter.
%--------------------------------------------------------------------%
The above systematic uncertainties do not scale with the event yield
and are included in the calculation of the significance of the result.
%--------------------------------------------------------------------%

%--------------------------------------------------------------------%
We estimate the systematic uncertainty due to the interference between 
the $B \rightarrow K_1 \pi$ and the $B \ra K^{*}\pi\pi + \rho 
K \pi$ decays using simulated samples in which the decay amplitudes 
are generated according to the results of the likelihood scans. 
The overall 
phases and relative contribution for the $K^{*}\pi\pi$ and $\rho K \pi$
interfering states are assumed to be constant across phase space
and varied between zero and a maximum value using uniform prior 
distributions. We calculate the systematic uncertainty from the RMS 
variation of the average signal branching fraction and parameters.
This uncertainty is assumed to scale as the square root of the 
signal branching fraction and does not affect the significance.
%--------------------------------------------------------------------%
The systematic uncertainties in efficiencies include
those associated with  track finding, particle identification and,
for the $B^+$ modes, $K_S^0$ reconstruction.
%--------------------------------------------------------------------%
Other systematic effects arise from event selection criteria,
such as track multiplicity and thrust angle,
and the number of $B$ mesons.        
%--------------------------------------------------------------------%

\begin{table}[h]   
\begin{center}
\caption{Estimates of systematic errors, evaluated at the absolute minimum
of each $-\ln{\cal L}$ scan. 
For the branching fraction, the errors labeled (A), for additive, 
are given in units of $10^{-6}$, while those labeled (M), for multiplicative,
are given in percentage.} 
\label{tab:systtab}
\small
\begin{tabular}{l|ccc|cc}
\hline\hline
         & \multicolumn{3}{c|}{$B^0 \ra K_1^+ \pi^-$ }& \multicolumn{2}{c}{$B^+ \ra K_1^0 \pi^+$}\\
\hline
Quantity & ${\cal B}$ & $\vartheta$ & $\phi$ & ${\cal B}$ & $\vartheta$ \\
\hline
%--------------------------------------------------------------------%
PDF parameters (A)     & $ 0.8 $ & $ 0.01 $ & $ 0.15 $  & $ \phantom{1}1.4  $ & $ 0.07 $ \\

MC/data correction (A) & $ 0.8 $ & $ 0.00 $ & $ 0.01 $  & $  \phantom{1}1.0  $ & $ 0.02 $ \\
%--------------------------------------------------------------------%
ML fit bias   (A)    & $0.6$  & $ 0.03 $ & $ 0.02 $ & $  \phantom{1}2.0  $ & $ 0.08 $  \\

Fixed phase (A)      & $-$    & $-$      & $-$      & $ \phantom{1}0.6$    & $ 0.06$    \\

Scan   (A)           & $0.9 $ & $0.04$  &  $0.16$  & $  \phantom{1}0.0  $ & $ 0.04 $  \\
%--------------------------------------------------------------------%
$K_1$ $K-$matrix parameters (A) & $ 2.2 $ & $ 0.01 $ & $0.36$ & $  \phantom{1}0.5  $ & $ 0.05 $  \\

$K_1$ offset phases (A) & $ 0.2 $ & $ 0.01 $ & $0.02$ & $  \phantom{1}0.0  $ & $ 0.00 $  \\

$K_1$ intermediate resonances (A) & $ 0.5 $ & $ 0.00 $ & $0.06$ & $  \phantom{1}0.2  $ & $ 0.02 $ \\
%--------------------------------------------------------------------%
$K^*/\rho$ bands (A) & $ 0.2 $   & $ 0.05 $ & $ 0.00 $  & $  \phantom{1}1.2  $ & $ 0.05 $  \\
%--------------------------------------------------------------------%
Peaking $\BB$ bkg (A) & $0.8$ & $ 0.01$ & $0.13 $ & $  \phantom{1}1.0  $ & $ 0.01 $  \\
%--------------------------------------------------------------------%
Fixed background yields (A) & $ 0.0 $   & $ 0.00 $ & $ 0.00 $  & $  \phantom{1}0.4  $ & $ 0.02 $  \\
%--------------------------------------------------------------------%

%--------------------------------------------------------------------%
Interference (A) & $ 6.0 $ & $ 0.25 $ & $ 0.52 $  & $ 10.6  $ & $ 0.43 $  \\
%--------------------------------------------------------------------%

MC statistics (M)     & $1.0$  & $-$  & $-$  & $  \phantom{1}1.0  $ & $ -    $  \\

Particle identification (M) & $2.9 $ & $-$  & $-$ & $  \phantom{1}3.1  $ & $ -    $  \\

Track finding (M)     & $1.0$  & $-$  & $-$ & $  \phantom{1}0.8  $ & $ -    $  \\

$K_S^0$ reconstruction (M) & $-  $  & $-$  & $-$ & $  \phantom{1}1.6 $ & $ -    $  \\
%--------------------------------------------------------------------%
\costhr       (M)     & $1.0 $ & $-$  & $-$ & $  \phantom{1}1.0  $ & $ -    $  \\

Track multiplicity  (M)    & $1.0$  & $-$  & $-$ & $  \phantom{1}1.0  $ & $ -    $  \\

Number \BB\ pairs (M)       & $1.1 $ & $-$  & $-$ & $  \phantom{1}1.1  $ & $ -    $  \\ 
%--------------------------------------------------------------------%

\hline\hline
\end{tabular}
\end{center}
\end{table}

\section{Fit results}
\label{sec:results}
\begin{figure*}[p]
  \begin{center}
    \includegraphics[width=6.5in]{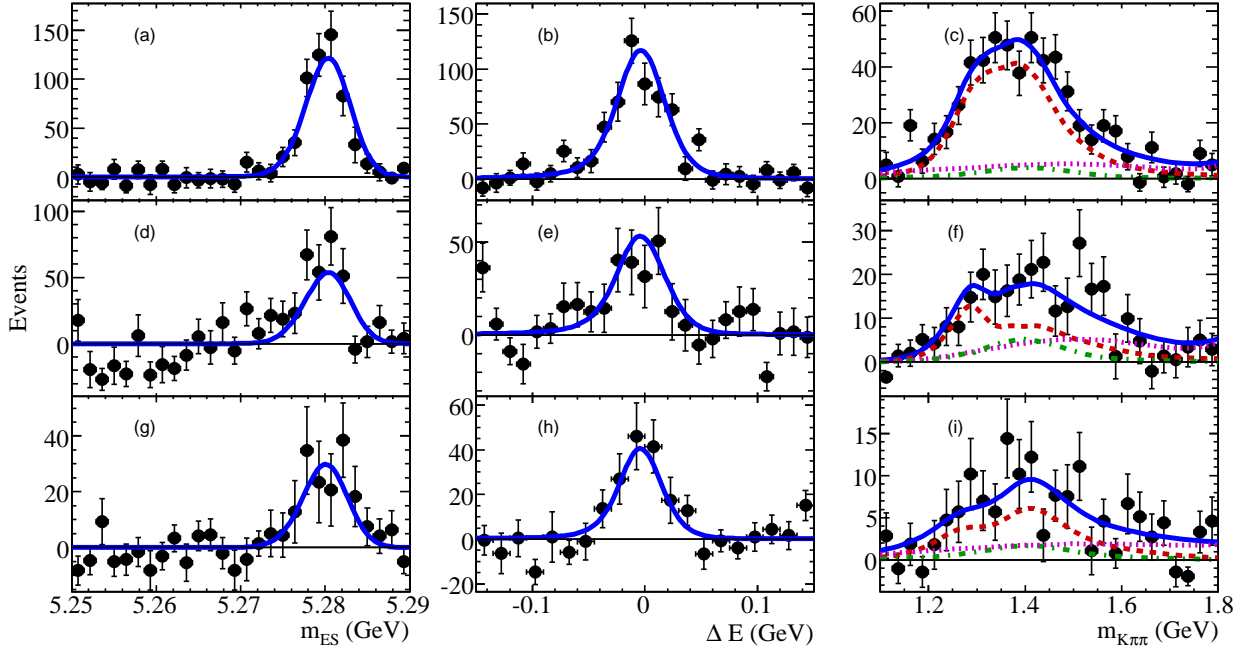}
 \caption{
   sPlot projections of signal onto \mes\ (left), $\Delta E$\ (center), 
   and $m_{K \pi \pi}$ 
   (right) for $B^0$ class 1 (top), $B^0$ class 2 (middle), and $B^+$ class 1 
   (bottom) events: 
   the points show the sums of the signal weights obtained from on-resonance data.
   For \mes\ and $\Delta E$ the solid line is the signal fit function.
   For $m_{K \pi \pi}$ the solid line is
   the sum of the fit functions of the
   decay modes $K_1(1270) \pi + K_1(1400) \pi$ (dashed), 
   $K^*(1410) \pi$ (dash-dotted), and $K^*(892) \pi \pi$ (dotted), and
   the points are obtained without using information about
   resonances in the fit, \emph{i.e.}, we use only the \mes, 
   $\Delta E$, and \xf\ variables.
 } 
 \label{fig:splots}
  \end{center}
\end{figure*}

\begin{figure*}[p]
  \begin{center}
    \includegraphics[width=6.5in]{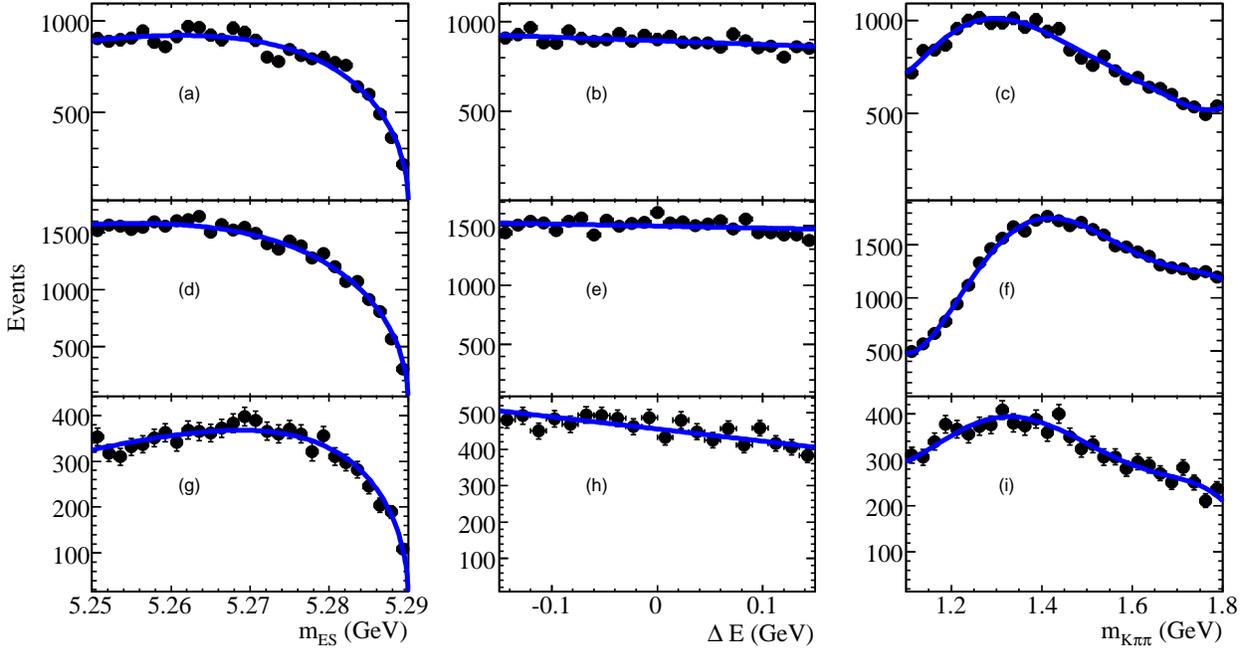}
 \caption{
   sPlot projections of combinatorial background onto 
   \mes\ (left), $\Delta E$\ (center), and 
   $m_{K \pi \pi}$ 
   (right) for $B^0$ class 1 (top), $B^0$ class 2 (middle), and $B^+$ class 1 
   (bottom) events: 
   the points show the sums of the combinatorial  background weights 
   obtained from on-resonance data.
   The solid line is the combinatorial background fit function.
   For $m_{K \pi \pi}$ 
   the points are obtained without using information about
   resonances in the fit, \emph{i.e.}, we use only the \mes, 
   $\Delta E$, and \xf\ variables.
 } 
 \label{fig:splotsqq}
  \end{center}
\end{figure*}

Figures~\ref{fig:splots} and \ref{fig:splotsqq} 
show the distributions of $\Delta E$, \mes\, and  
$m_{K\pi\pi}$ for the signal and combinatorial background events, respectively,
obtained by the event-weighting technique $_s\cal{P}$lot~\cite{SPLOT}. 
For each event, signal and background weights are derived according
to the results of the fit to all variables and the probability distributions 
in the restricted set of variables in which the projection variable is
omitted. Using these weights, the data are then plotted as a function of the projection 
variable.

\begin{figure*}[!htb]
\begin{center}
  \begin{minipage}{3in}
    \centering
    \includegraphics[width=3.in]{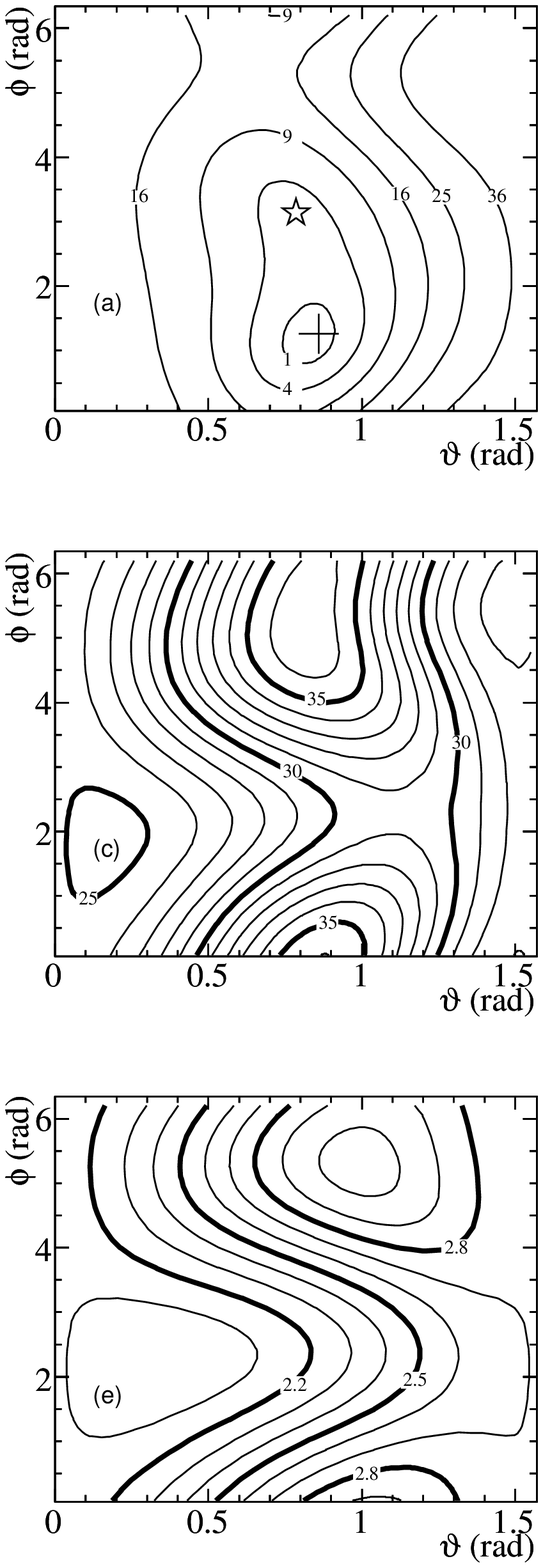}
  \end{minipage}
  \begin{minipage}{3in}
    \centering
    \includegraphics[width=3.in]{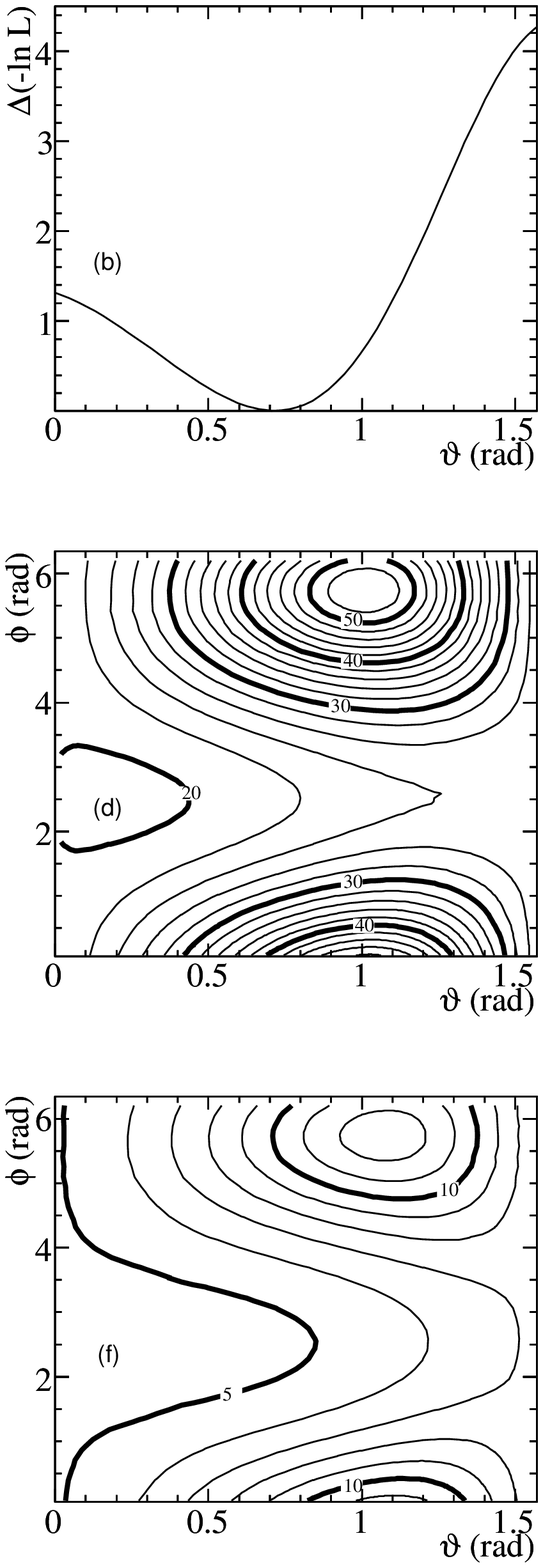}
  \end{minipage}
  \caption{(a, b) $-\ln{\cal L}$ scan (systematics not included) 
    in the production parameters 
    $\vartheta$ and $\phi$ for the (a) $B^0$ and (b) $B^+$ modes.
    The cross in (a) indicates the position of the absolute minimum in the
    $-\ln{\cal L}$ scan. A second, 
    local minimum is indicated 
    by a star and corresponds to an increase in $\Delta(-\ln{\cal L})$
    of $2.7$ 
    with respect to the absolute minimum.
    (c, d) Contours for the $B\to K_1(1270)\pi + K_1(1400)\pi$ branching 
    fraction (in units of $10^{-6}$) extracted from the ML fit for the 
    (c) $B^0$ and (d) $B^+$ modes.
    (e, f) Contours for the statistical error (in units of $10^{-6}$) on the 
    $B\to K_1(1270)\pi + K_1(1400)\pi$ branching 
    fraction for the (e) $B^0$ and (f) $B^+$ modes.
  }
  \label{fig:nllscan}
\end{center}
\end{figure*}

\begin{figure*}[h]
\begin{center}
  \begin{minipage}{3in}
    \centering
    \includegraphics[width=3.in]{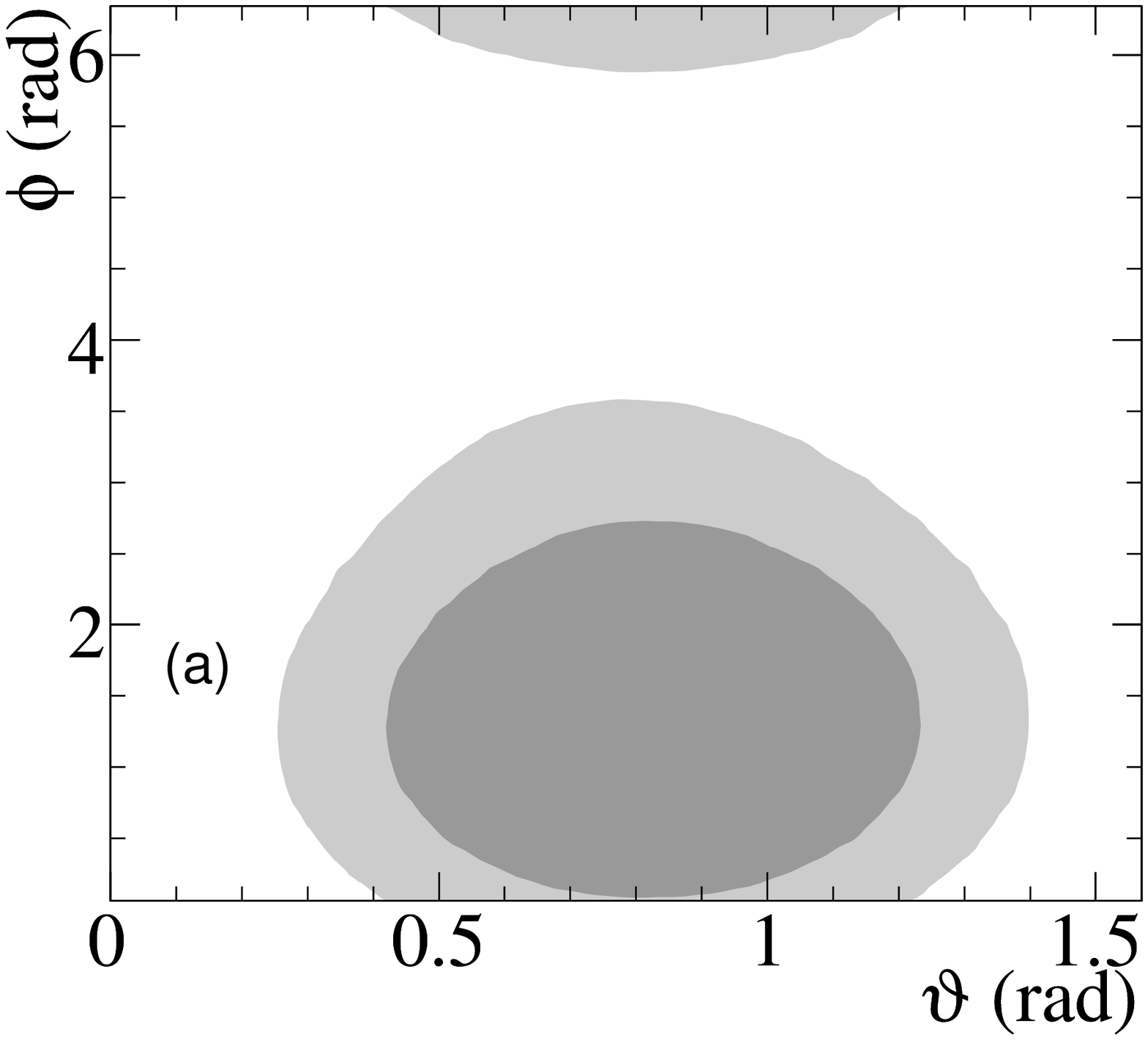}
  \end{minipage}
  \begin{minipage}{3in}
    \centering
    \includegraphics[width=3.in]{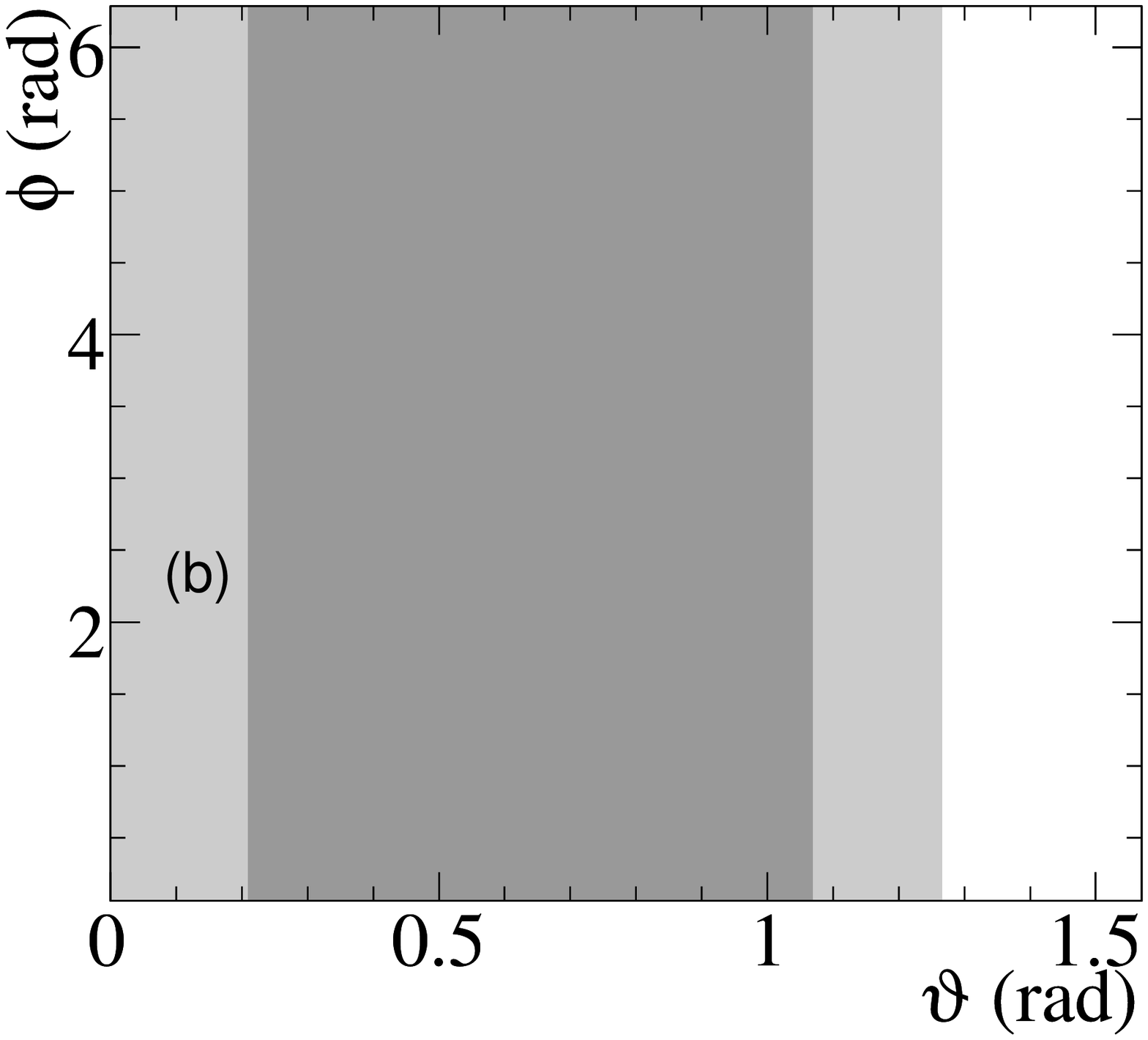}
  \end{minipage}
  \caption{(a, b) 68\% (dark shaded zone) and 90\% (light shaded zone) 
    probability regions for $\vartheta$ and $\phi$ for the (a) $B^0$ 
    and (b) $B^+$ modes.
  }
  \label{fig:probscan}
\end{center}
\end{figure*}

\begin{table}[b]   
\begin{center}
\caption{Results of the ML fit at the  absolute minimum of the
$-\ln{\cal L}$ scan.
The first two rows report the values of the production parameters 
$(\vartheta,\phi)$ that maximize the likelihood. The third and fourth
rows are the reconstruction efficiencies, including the daughter branching 
fractions, for class 1 and class 2 events. The fifth row is the
correction for the fit bias to the signal branching fraction. The sixth 
row reports the results for the $B\to K_1(1270)\pi + K_1(1400)\pi$
branching fraction and its error (statistical only).}
\label{tab:restab}
\small
\begin{tabular}{l|c|c}
\hline\hline
 &$B^0 \ra K_1^+\pi^-$ &  $B^+ \ra K_1^0 \pi^+$\\
\hline
$\vartheta$  & $0.86$ & $0.71$ \\
$\phi$       & $1.26$ & $3.14$ (fixed) \\
$\epsilon_1$ (\%)& $3.74$   & $1.36$  \\
$\epsilon_2$ (\%)& $1.68$   & --    \\
Fit bias correction  $(\times 10^{-6})$ & $+0.0$ & $+0.7$ \\
${\cal B}$ $(\times 10^{-6})$           & $32.1\pm 2.4$     & $22.8 \pm 5.1$ \\
\hline\hline
\end{tabular}
\end{center}
\end{table}   
The results of the likelihood scans are shown in Table~\ref{tab:restab}
and Fig.~\ref{fig:nllscan}. 
At each point of the  ${\boldsymbol \zeta}$ scan the 
$-2\ln {\cal L}({\cal B};{\boldsymbol \zeta})$ function is minimized 
with respect to the signal branching fraction ${\cal B}$.
Contours for the value ${\cal B}_{\rm max}({\boldsymbol \zeta})$ that 
maximizes ${\cal L}({\cal B};{\boldsymbol \zeta})$ are shown 
in Fig.~\ref{fig:nllscan}c and Fig.~\ref{fig:nllscan}d 
as a function of the production parameters ${\boldsymbol \zeta}$, for 
$B^0$ and $B^+$ modes, respectively.
The associated statistical error $\sigma_{\cal B}({\boldsymbol \zeta})$ 
at each point ${\boldsymbol \zeta}$, 
given by the change in ${\cal B}$ when the quantity 
$-2\ln {\cal L}({\cal B};{\boldsymbol \zeta})$ increases by one unit, 
is displayed in Fig.~\ref{fig:nllscan}e and Fig.~\ref{fig:nllscan}f.
Systematics are included by convolving the experimental 
two-dimensional likelihood for $\vartheta$ and $\phi$,
$\mathcal{L}\equiv{\cal L}({\cal B}_{\rm max}({\boldsymbol \zeta});{\boldsymbol \zeta})$, 
with a two-dimensional Gaussian that accounts for the systematic
uncertainties.
In Fig.~\ref{fig:probscan}a and Fig.~\ref{fig:probscan}b  
we show the resulting distributions in 
$\vartheta$ and $\phi$. The 68\% and 90\% probability 
regions are shown in dark and light shading, respectively, and
are defined as the regions consisting of all the points that 
satisfy the condition ${\cal {L}}(r)>x$, where the value $x$ is such that  
$\int_{{\cal {L}}(r)>x}{\cal {L}}(\vartheta,\phi) {\rm d}\vartheta{\rm d}\phi = 68\%~~(90\%)$. 
The significance is calculated from a likelihood ratio test
$ \Delta(-2\ln{\cal L}) $ evaluated at the value of $\vartheta$
that maximizes the likelihood averaged over $\phi$. 
Here $ \Delta(-2\ln{\cal L}) $ is the 
difference between the value of $-2\ln{\cal L}$ (convolved with systematic 
uncertainties) for zero signal and the value at its minimum for
given values of ${\boldsymbol \zeta}$.
We calculate the significance from a $\chi^2$ distribution for 
$\Delta(-2\ln{\cal L})$ with 2 degrees of freedom. 
We observe nonzero $B^0 \ra K_1^+\pi^-$ and $B^+ \ra K_1^0 \pi^+$ 
branching fractions with $7.5\sigma$ and $3.2\sigma$ significance,
respectively.

\begin{table*}[h]   
\begin{center}
\caption{Branching fraction results for $B\to K_1\pi$ decays, 
in units of $10^{-5}$, and corresponding confidence levels 
(C.L., statistical uncertainties only). 
For each branching fraction we provide the mean 
of the probability distribution, the most probable value (MPV), 
the two-sided interval at 68\% probability, and the 
upper limit at 90\% probability.}
\label{tab:BFstat}
\begin{tabular}{l|c|c|c|c}
\hline\hline
Channel & Mean & MPV & 68\% C.L. interval & 90\% C.L. UL \\
\hline
$B^0 \to K_1(1270)^+\pi^- + K_1(1400)^+\pi^-$ & $3.2 $ & $3.1$ & $(2.9,3.4)$ & $3.5$ \\
$B^0 \to K_1(1270)^+\pi^-$ & $1.7$ & $1.6$ & $(1.3,2.0)$ & $2.1$ \\
$B^0 \to K_1(1400)^+\pi^-$ & $1.6$ & $1.6$ & $(1.3,1.9)$ & $2.0$ \\
$B^0 \to K_{1A}^+\pi^-$ & $1.5$ & $1.4$ & $(1.0,1.9)$ & $2.2$ \\
\hline
$B^+ \to K_1(1270)^0\pi^+ + K_1(1400)^0\pi^+$ & $2.9 $ & $2.3$ & $(1.6,3.5)$ & $4.5$ \\
$B^+ \to K_1(1270)^0\pi^+$ &  $1.1$ & $0.3$ & $(0.0,1.4)$ & $2.5$ \\
$B^+ \to K_1(1400)^0\pi^+$ & $1.8$ & $1.7$ & $(1.0,2.5)$ & $2.0$ \\
$B^+ \to K_{1A}^0\pi^+$ & $1.1$ & $0.2$ & $(0.0,1.5)$ & $2.3$ \\
\hline\hline
\end{tabular}
\end{center}
\end{table*}

\begin{table*}[h]   
\begin{center}
\caption{Branching fraction results for $B\to K_1\pi$ decays, 
in units of $10^{-5}$, and corresponding confidence levels 
(C.L., systematic uncertainties included). 
For each branching fraction 
we provide the mean 
of the probability distribution, the most probable value (MPV), 
the two-sided interval at 68\% probability, and the 
upper limit at 90\% probability.}
\label{tab:BFcombined}
\begin{tabular}{l|c|c|c|c}
\hline\hline
Channel & Mean & MPV & 68\% C.L. interval & 90\% C.L. UL \\
\hline
$B^0 \to K_1(1270)^+\pi^- + K_1(1400)^+\pi^-$ & $3.3 $ & $3.1$ & $(2.4,3.9)$ & $4.3$ \\ 
$B^0 \to K_1(1270)^+\pi^-$ & $1.7$ & $1.7$ & $(0.6,2.5)$ & $3.0$ \\ 
$B^0 \to K_1(1400)^+\pi^-$ & $1.6$ & $1.7$ & $(0.8,2.4)$ & $2.7$ \\ 
$B^0 \to K_{1A}^+\pi^-$ & $1.6$ & $1.4$ & $(0.4,2.3)$ & $2.9$ \\ 
\hline
$B^+ \to K_1(1270)^0\pi^+ + K_1(1400)^0\pi^+$ & $4.6 $ & $2.9$ & $(1.2,5.8)$ & $8.2$ \\ 
$B^+ \to K_1(1270)^0\pi^+$ & $1.7$ & $0.0$ & $(0.0,2.1)$ & $4.0$ \\ 
$B^+ \to K_1(1400)^0\pi^+$ & $2.0$ & $1.6$ & $(0.0,2.5)$ & $3.9$ \\
$B^+ \to K_{1A}^0\pi^+$ & $1.6$ & $0.2$ & $(0.0,2.1)$ & $3.6$ \\ 
\hline\hline
\end{tabular}
\end{center}
\end{table*}

We derive probability distributions for the 
$B \to K_1 (1270) \pi +K_1(1400) \pi$, 
$B\to K_1(1270)\pi$, 
$B\to K_1(1400)\pi$, and 
$B\to K_{1A}\pi$  
branching fractions.

At each point in the ${\boldsymbol \zeta}$ plane we calculate 
the distributions for the branching fractions, given by
$f({\cal B};{\boldsymbol \zeta}) = c {\cal {L}}({\cal B};{\boldsymbol \zeta})$, 
where $c$ is a normalization
constant. Systematics are included by convolving the experimental 
one-dimensional likelihood 
$\mathcal{L}({\cal B};{\boldsymbol \zeta})$ with a Gaussian 
that represents systematic uncertainties.
Branching fraction results are obtained by means of a weighted average
of the branching fraction distributions defined above, with
weights calculated from the experimental two-dimensional likelihood 
for $\vartheta$ and $\phi$.

For each point of the ${\boldsymbol \zeta}$ scan the $B\to K_1(1270)\pi$, 
$B\to K_1(1400)\pi$, and $B\to K_{1A}\pi$ branching fractions are 
obtained by applying ${\boldsymbol \zeta}$-dependent correction
factors to the $B \to K_1(1270) \pi +K_1(1400) \pi$ branching fraction 
associated with that ${\boldsymbol \zeta}$ point.
The correction factor is calculated by reweighting the signal MC samples  
by setting the production parameters $(f_{pa},f_{pb})$ equal to 
$(0,e^{i\phi}\sin\vartheta)$, $(\cos\vartheta,0)$, and 
$(|f_{pA}|\cos\theta,-|f_{pA}|\sin\theta)$, for $B\to K_1(1270)\pi$, 
$B\to K_1(1400)\pi$, and $B\to K_{1A}\pi$, respectively,
where $f_{pA}=\cos\vartheta\cos\theta - e^{i\phi}\sin\vartheta \sin\theta$ and
$\theta$ is the $K_1$ mixing angle \cite{WA3}, for which we use the 
value $\theta = 72^{\circ}$ (see Table~\ref{tab:daumfit}).

From the resulting distributions $f({\cal B})$ we calculate the 
corresponding two-sided intervals at 68\% probability, which consist of 
all the points ${\cal B}>0$ that satisfy the condition
$f({\cal B})>x$, where $x$ is such that  
$\int_{f({\cal B})>x,~{\cal B}>0}f({\cal B}) {\rm d}{\cal B}= 68\%$. 
The upper limits (UL) at 90\% probability are calculated as
$\int_{0<{\cal B}<UL}f({\cal B}) {\rm d}{\cal B}= 90\%$. 
The results are summarized in Table~\ref{tab:BFstat} (statistical only) 
and Table~\ref{tab:BFcombined} (including systematics).

We measure
${\cal B}(B^0\to K_1(1270)^{+}\pi^-+K_1(1400)^{+}\pi^-) = 3.1^{+0.8}_{-0.7} \times 10^{-5}$ and  
${\cal B}(B^+\to K_1(1270)^{0}\pi^++K_1(1400)^{0}\pi^+) = 2.9^{+2.9}_{-1.7} \times 10^{-5}$ ($<8.2\times 10^{-5}$), where the two-sided ranges and upper limits are 
evaluated at 68\% and 90\% probability, respectively, and include systematic 
uncertainties.

Including systematic uncertainties we obtain the two-sided  
intervals (in units of $10^{-5}$):
${\cal B}(B^0 \to K_1(1270)^{+}\pi^-) \in [0.6,2.5]$,~
${\cal B}(B^0 \to K_1(1400)^{+}\pi^-) \in [0.8,2.4]$,~
${\cal B}(B^0 \to K_{1A}^{+}\pi^-) \in [0.4,2.3]$,~
${\cal B}(B^+ \to K_1(1270)^{0}\pi^+) \in [0.0,2.1]$~($<4.0$),~
${\cal B}(B^+ \to K_1(1400)^{0}\pi^+) \in [0.0,2.5)$~($<3.9$),~
${\cal B}(B^+ \to K_{1A}^{0}\pi^+) \in [0.0,2.1]$~($<3.6$),~
where the two-sided ranges and the upper limits are 
evaluated at 68\% and 90\% probability, respectively.

\section{Bounds on $\boldsymbol {|\Delta \alpha|}$}
\label{sec:alpha}
We use the measurements
presented in this work to derive 
bounds on the model uncertainty
$|\Delta \alpha|$ on the weak phase $\alpha$ extracted in 
$B^0\rightarrow a_1(1260)^{\pm}\pi^{\mp}$ decays.
We use the previously measured branching fractions of 
$B^0\rightarrow a_1(1260)^{\pm}\pi^{\mp}$, 
$B^0\rightarrow a_1(1260)^{-}K^{+}$ and 
$B^+\rightarrow a_1(1260)^{+}K^0$ decays \cite{APDECAYSA1}
and the $CP-$violation asymmetries \cite{ALPHA} 
as input to the method of Ref.~\cite{BOUNDS}.
The values used are summarized in Tables~\ref{tab:alphabfinput} and~\ref{tab:alphacpinput}.

\begin{table}[h]
\caption{Summary of the branching fractions used as input to the 
calculation of the bounds on $|\Delta \alpha|$~\cite{APDECAYSA1}.}
\label{tab:alphabfinput}
\begin{center}
\begin{tabular}{l|c}
\hline\hline
Decay mode & Branching fraction\\
&  (in units of $10^{-6}$) \\ 
\hline
$B^0\to a_1(1260)^{\pm}\pi^{\mp} $ &  $33.2\pm 3.8 \pm 3.0$\\
$B^0\to a_1(1260)^-K^+ $ & $16.3\pm 2.9 \pm 2.3$\\
$B^+\to a_1(1260)^+K^0 $ & $33.2\pm 5.0 \pm 4.4$\\
\hline\hline
\end{tabular}
\end{center}
\end{table}

\begin{table}[h]
\caption{Summary of the values of the $CP-$violation parameters 
used as input to the calculation of the bounds on 
$|\Delta \alpha|$~\cite{ALPHA}.}
\label{tab:alphacpinput}
\begin{center}
\begin{tabular}{l|c}
\hline\hline
Quantity & Value \\
\hline 
${\cal A}_{CP}^{a_1\pi} $ &$-0.07\pm 0.07\pm 0.02 $ \\
$S $             &$ \phantom{-}0.37 \pm 0.21 \pm 0.07$ \\
$\Delta S $      &$ -0.14 \pm 0.21 \pm 0.06$ \\
$C $             &$ -0.10 \pm 0.15 \pm 0.09$ \\
$\Delta C $      &$ \phantom{-}0.26 \pm 0.15 \pm 0.07$ \\
\hline\hline
\end{tabular}
\end{center}
\end{table}

The bounds are calculated as the average of 
$|\Delta \alpha|^+ = |\alpha_{\rm eff}^+-\alpha|$ and
$|\Delta \alpha|^- = |\alpha_{\rm eff}^--\alpha|$, which 
are obtained from the inversion of the relations~\cite{BOUNDS}:
\begin{eqnarray}
\cos2(\alpha_{\rm eff}^{\pm}-\alpha)\geq \frac{1-2 R^{0}_{\pm}}{\sqrt{1-{\cal A}_{CP}^{\pm~2}}},\nonumber \\
\cos2(\alpha_{\rm eff}^{\pm}-\alpha)\geq \frac{1-2 R^{+}_{\pm}}{\sqrt{1-{\cal A}_{CP}^{\pm~2}}},
\label{cosdeltaalpha}
\end{eqnarray}
where we have defined the following ratios of $CP$-averaged rates~\cite{BOUNDS}:
\begin{eqnarray*}
R^0_+ &\equiv &\frac{\bar{\lambda}^2f_{a_1}^2\bar{\cal B}(K_{1A}^+\pi^-)}{f_{K_{1A}}^2\bar{\cal B}(a_1^+\pi^-)}\\
R^0_- &\equiv &\frac{\bar{\lambda}^2f_{\pi}^2\bar{\cal B}(a_1^-K^+)}{f_{K}^2\bar{\cal B}(a_1^-\pi^+)}\\
R^+_+ &\equiv &\frac{\bar{\lambda}^2f_{a_1}^2\bar{\cal B}(K_{1A}^0\pi^+)}{f_{K_{1A}}^2\bar{\cal B}(a_1^+\pi^-)}\\
R^+_- &\equiv &\frac{\bar{\lambda}^2f_{\pi}^2\bar{\cal B}(a_1^+K^0)}{f_{K}^2\bar{\cal B}(a_1^-\pi^+)}.
\end{eqnarray*}

The $CP$ asymmetries ${\cal A}_{CP}^{\pm}$ in 
$B^0 \rightarrow a_1^{\pm}\pi^{\mp}$ decays are related to the time- and 
flavor-integrated charge asymmetry ${\cal A}_{CP}^{a_1\pi}$ \cite{ALPHA}
by
%%%
\begin{eqnarray*}
{\cal A}_{CP}^+&=&-\frac{{\cal A}_{CP}^{a_1\pi}(1+\Delta C)+C}{1+{\cal A}_{CP}^{a_1\pi}C+\Delta C},\\
{\cal A}_{CP}^-&=&\frac{{\cal A}_{CP}^{a_1\pi}(1-\Delta C)-C}{1-{\cal A}_{CP}^{a_1\pi}C-\Delta C}.
\end{eqnarray*}
$C$ and $\Delta C$ parameterize the flavor-dependent direct $CP$ violation
and the asymmetry between the $CP$-averaged rates $\bar{\cal B}(a_1^+\pi^-)$ 
and $\bar{\cal B}(a_1^-\pi^+)$, respectively~\cite{BOUNDS}:
\begin{equation*}
C \pm \Delta C \equiv \frac{|A_{\pm}|^2 - |\overline{A}_{\mp}|^2}{|A_{\pm}|^2 + 
|\overline{A}_{\mp}|^2}~,
\end{equation*}
where the decay amplitudes for 
$B^0(\overline B^0)\to a_1(1260)^{\pm}\pi^{\mp}$ are 
\begin{eqnarray*}
A_+ & \equiv & A(B^0\to a_1^+\pi^-)~,~~~ 
A_-  \equiv A(B^0\to a_1^-\pi^+)~,\nonumber\\
\overline{A}_+ & \equiv & A({\overline B}^0 \to a_1^-\pi^+)~,~~~ 
\overline{A}_- \equiv A({\overline B}^0 \to a_1^+\pi^-)~.
\end{eqnarray*}

The $CP$-averaged rates are calculated as
\begin{eqnarray*}
\bar{\cal B}(a_1^+\pi^-)=\frac{1}{2}{\cal B}(a_1^{\pm}\pi^{\mp})(1+\Delta C + {\cal A}_{CP}^{a_1\pi}C),\\
\bar{\cal B}(a_1^-\pi^+)=\frac{1}{2}{\cal B}(a_1^{\pm}\pi^{\mp})(1-\Delta C - {\cal A}_{CP}^{a_1\pi}C),
\end{eqnarray*}
where ${\cal B}(a_1^{\pm}\pi^{\mp})$ is the flavor-averaged branching fraction
of neutral $B$ decays to $a_1(1260)^{\pm}\pi^{\mp}$ \cite{APDECAYSA1}.

For the constant $\bar{\lambda}=|V_{us}|/|V_{ud}| = |V_{cd}|/|V_{cs}|$ we
take the value 0.23 \cite{PDG}. The decay constants $f_{K}=155.5 \pm 0.9$~\mev and
$f_{\pi}=130.4 \pm 0.2$~\mev\ \cite{PDG} are experimentally known with small 
uncertainties. For the decay constants of the $a_{1}$ and $K_{1A}$ mesons
the values $f_{a_{1}}=203 \pm 18$~\mev\ \cite{BLOCH} and 
$f_{K_{1A}}=207$~\mev\ \cite{CHENG} are used. For $f_{K_{1A}}$ we assume an
uncertainty of $20$~\mev.
The value assumed for the $f_{K_{1A}}$ decay constant 
is based on a mixing angle 
$\theta = 58^{\circ}$ \cite{CHENG}, because $f_{K_{1A}}$
is not available for the value $\theta = 72^{\circ}$ used 
here (see Table~\ref{tab:daumfit});
this discrepancy is likely 
accommodated within the accuracy of the present experimental 
constraints on the mixing angle.
Using na\"{\i}ve arguments based on SU(3) relations and the mixing formulae,
we have verified that the dependence of  $f_{K_{1A}}$ on the mixing angle is
rather mild in the $\theta$ range $[58,72]^{\circ}$.
It should be noted that due to a different choice of notation, a positive 
mixing angle in the formalism used by the ACCMOR Collaboration \cite{WA3} and 
in this paper corresponds to a negative mixing angle with the notation of 
Ref.~\cite{CHENG}.

We use a Monte Carlo technique to estimate a probability region for the 
bound on $|\alpha_{\rm eff}-\alpha|$. All the $CP$-averaged rates 
and $CP$-violation parameters participating in 
the estimation of the bound are generated according to the experimental 
distributions, taking into account the statistical correlations among 
${\cal A}_{CP}^{a_1\pi}$, $C$, and $\Delta C$~\cite{HFAG}.

For each set of generated values we solve the system of inequalities in 
Eq. (\ref{cosdeltaalpha}), which involve $|\alpha_{\rm eff}^+-\alpha|$ and 
$|\alpha_{\rm eff}^--\alpha|$, and calculate the bound on $|\alpha_{\rm eff}-\alpha|$
from 
\begin{equation}
|\alpha_{\rm eff}-\alpha|\le (|\alpha_{\rm eff}^+-\alpha|+|\alpha_{\rm eff}^--\alpha|)/2.
\end{equation}
The probability regions are obtained by a counting method: we estimate the
fraction of experiments with a value of the bound on $|\alpha_{\rm eff}-\alpha|$ 
greater than a given value. 
We obtain $|\alpha_{\rm eff}-\alpha|<11^{\circ}(13^{\circ})$ at 68\% (90\%) probability.

The determination of $\alpha_{\rm eff}$ \cite{ALPHA} presents an eightfold ambiguity
in the range $[0^{\circ},180^{\circ}]$. The eight solutions are 
$\alpha_{\rm eff}=(11 \pm 7)^{\circ}$, $\alpha_{\rm eff}=(41 \pm 7)^{\circ}$,
$\alpha_{\rm eff}=(49 \pm 7)^{\circ}$, $\alpha_{\rm eff}=(79 \pm 7)^{\circ}$, 
$\alpha_{\rm eff}=(101 \pm 7)^{\circ}$, $\alpha_{\rm eff}=(131 \pm 7)^{\circ}$,
$\alpha_{\rm eff}=(139 \pm 7)^{\circ}$, $\alpha_{\rm eff}=(169 \pm 7)^{\circ}$~\cite{ALPHA}.
Assuming that the relative strong phase between the relevant tree amplitudes is 
negligible \cite{BOUNDS} it is possible to reduce this ambiguity
to a twofold ambiguity in the range $[0^{\circ},180^{\circ}]$:
$\alpha_{\rm eff}=(11 \pm 7)^{\circ}$, $\alpha_{\rm eff}=(79 \pm 7)^{\circ}$.
We combine the solution near $90^{\circ}$, 
$\alpha_{\rm eff}=(79 \pm 7)^{\circ}$ \cite{ALPHA},
with the bounds on $|\alpha_{\rm eff}-\alpha|$ and estimate the 
weak phase 
$\alpha=(79 \pm 7 \pm 11)^{\circ}$. This solution is consistent 
with the current average value of $\alpha$, based on the 
analysis of $B\to \pi\pi$, $B\to \rho\rho$, and $B\to \rho\pi$ 
decays \cite{PDG,AVERAGES}.

\section{Summary}

We present results from a branching fraction measurement of 
$B\to K_1(1270)\pi$ and $K_1(1400)\pi$ decays, obtained from a data
sample of 454 million $\Upsilon(4S)\to B\bar{B}$ events.
The signal is modeled with a $K$-matrix formalism, which 
accounts for the effects of interference between the $K_1(1270)$ 
and $K_1(1400)$ mesons.
Including systematic and model uncertainties, we measure 
${\cal B}(B^0\to K_1(1270)^{+}\pi^-+K_1(1400)^{+}\pi^-)= 3.1^{+0.8}_{-0.7} \times 10^{-5}$ and 
${\cal B}(B^+\to K_1(1270)^{0}\pi^++K_1(1400)^{0}\pi^+)= 2.9^{+2.9}_{-1.7} \times 10^{-5}$ ($<8.2\times 10^{-5}$ at 90\% probability).
A combined signal for the decays $ B^0\to K_1(1270)^{+}\pi^-$ and $B^0\to K_1(1400)^{+}\pi^-$ is observed with
a significance of $7.5\sigma$, and the following branching fractions 
are derived for neutral $B$ meson decays:
${\cal B}(B^0 \to K_1(1270)^{+}\pi^-) \in [0.6,2.5]\times 10^{-5}$,~
${\cal B}(B^0 \to K_1(1400)^{+}\pi^-) \in [0.8,2.4]\times 10^{-5}$, and 
${\cal B}(B^0 \to K_{1A}^{+}\pi^-) \in [0.4,2.3]\times 10^{-5}$,~
where the two-sided intervals are evaluated at 68\% probability.
A significance of $3.2\sigma$
is obtained for $B^+\to K_1(1270)^{0}\pi^++K_1(1400)^{0}\pi^+$,
and we derive the following two-sided intervals at 68\% probability and 
upper limits at 90\% probability:
${\cal B}(B^+ \to K_1(1270)^{0}\pi^+) \in [0.0,2.1]\times 10^{-5}$~($<4.0\times 10^{-5}$),~
${\cal B}(B^+ \to K_1(1400)^{0}\pi^+) \in [0.0,2.5)\times 10^{-5}$~($<3.9\times 10^{-5}$), and 
${\cal B}(B^+ \to K_{1A}^{0}\pi^+) \in [0.0,2.1]\times 10^{-5}$~($<3.6\times 10^{-5}$).

Finally, we combine the results presented in this paper with
existing experimental information to derive an independent 
estimate for the CKM angle $\alpha$, 
based on the time-dependent analysis of $CP$-violating asymmetries in 
$B^0 \ra a_1(1260)^{\pm} \pi^{\mp}$, and find $\alpha=(79 \pm 7 \pm 11)^{\circ}$.

\section{Acknowledgements}

We thank Ian Aitchison for helpful discussions and suggestions.~We are grateful for the 
extraordinary contributions of our \pep2\ colleagues in
achieving the excellent luminosity and machine conditions
that have made this work possible.
The success of this project also relies critically on the 
expertise and dedication of the computing organizations that 
support \babar.
The collaborating institutions wish to thank 
SLAC for its support and the kind hospitality extended to them. 
This work is supported by the
US Department of Energy
and National Science Foundation, the
Natural Sciences and Engineering Research Council (Canada),
the Commissariat \`a l'Energie Atomique and
Institut National de Physique Nucl\'eaire et de Physique des Particules
(France), the
Bundesministerium f\"ur Bildung und Forschung and
Deutsche Forschungsgemeinschaft
(Germany), the
Istituto Nazionale di Fisica Nucleare (Italy),
the Foundation for Fundamental Research on Matter (The Netherlands),
the Research Council of Norway, the
Ministry of Education and Science of the Russian Federation, 
Ministerio de Educaci\'on y Ciencia (Spain), and the
Science and Technology Facilities Council (United Kingdom).
Individuals have received support from 
the Marie-Curie IEF program (European Union) and
the A. P. Sloan Foundation.


\begin{thebibliography}{99}
\bibitem{CHARGE}
Except as noted explicitly, we use a particle name to denote either 
member of a charge conjugate pair.
\end{thebibliography}
\end{document}